\documentclass[fleqn,usenatbib]{mnras}
\usepackage{newtxtext,newtxmath}
\usepackage[T1]{fontenc}
\usepackage{ae,aecompl}

\usepackage{etoolbox}
\makeatletter
\patchcmd\@combinedblfloats{\box\@outputbox}{\unvbox\@outputbox}{}{\errmessage{\noexpand patch failed}}
\makeatother

\usepackage{graphicx}	
\usepackage{amsmath}	
\usepackage{amssymb}	
\usepackage{soul}
\usepackage{newtxtext}
\usepackage[english]{babel}
\usepackage{booktabs}
\usepackage{graphicx,color}
\usepackage{epstopdf}
\usepackage{mathtools}

\newcommand{\Mpch}{$h^{-1}\,\mbox{Mpc}$}

\newcommand{\xiis}{$\xi(s_\perp,s_\parallel)$~}

\newcommand{\ie}{{i.e.}}
\newcommand{\eg}{{e.g.}}

\newcommand{\mdpl}{{\small MDPL2} }
\newcommand{\lcdm}{$\mathrm{\Lambda CDM}$\,}

\definecolor{DR}{rgb}{0.9, 0.0, 0.1}

\title[Validation of RSD clustering modelling]{Validating the
  methodology for constraining the linear growth rate from
  clustering anisotropies} \author[Garc\'ia-Farieta J.E. et
  al. 2019]{\parbox{\textwidth}{Jorge Enrique
    Garc\'ia-Farieta$^{1,2}$\thanks{E-mail:
      \href{joegarciafa@unal.edu.co}{joegarciafa@unal.edu.co}},
    Federico Marulli$^{2,3,4}$, Lauro Moscardini$^{2,3,4}$, \\ Alfonso
    Veropalumbo$^{5,6}$, Rigoberto
    A. Casas-Miranda$^{1}$}\\ \\ $^1$Departamento de F\'isica,
  Universidad Nacional de Colombia - Sede Bogot\'a, Av. Cra 30 No
  45-03, Bogot\'a, Colombia\\ $^2$Dipartimento di Fisica e Astronomia,
  Alma Mater Studiorum Universit\`{a} di Bologna, via Gobetti 93/2,
  I-40129 Bologna, Italy\\ $^3$INAF - Osservatorio di Astrofisica e
  Scienza dello Spazio di Bologna, via Gobetti 93/3, I-40129 Bologna,
  Italy \\ $^4$INFN - Sezione di Bologna, viale Berti Pichat 6/2,
  I-40127 Bologna, Italy\\ $^5$Dipartimento di Fisica, Universit\`a
  degli Studi Roma Tre, via della Vasca Navale 84, I-00146 Rome,
  Italy\\ $^6$INFN - Sezione di Roma Tre, via della Vasca Navale 84,
  I-00146 Rome, Italy }
\date{Accepted XXX. Received YYY; in original form ZZZ}

\pubyear{2019}

\begin{document}
\label{firstpage}
\pagerange{\pageref{firstpage}--\pageref{lastpage}}
\maketitle

\begin{abstract}
Redshift-space clustering distortions provide one of the most powerful 
probes to test the gravity theory on the largest cosmological scales. 
We perform a systematic validation study of the state-of-the-art 
statistical methods currently used to constrain the linear growth rate 
from redshift-space distortions in the galaxy two-point correlation 
function. The numerical pipelines are tested on mock halo catalogues 
extracted from large N-body simulations of the standard cosmological 
framework. We consider both the monopole and quadrupole multipole 
moments of the redshift-space two-point correlation function, as well 
as the radial and transverse clustering wedges, in the comoving scale 
range $10<r[$\Mpch$]<55$. Moreover, we investigate the impact of 
redshift measurement errors on the growth rate and linear bias 
measurements due to the assumptions in the redshift-space distortion 
model. Considering both the dispersion model and two widely-used models 
based on perturbation theory, we find that the linear growth rate is 
underestimated by about $5-10\%$ at $z<1$, while limiting the analysis 
at larger scales, $r>30$ \Mpch, the discrepancy is reduced below $5\%$. 
At higher redshifts, we find instead an overall good agreement between 
measurements and model predictions. Though this accuracy is good enough 
for clustering analyses in current redshift surveys, the models have to 
be further improved not to introduce significant systematics in RSD 
constraints from next generation galaxy surveys. The effect of redshift 
errors is degenerate with the one of small-scale random motions, and 
can be marginalised over in the statistical analysis, not introducing 
any statistically significant bias in the linear growth constraints, 
especially at $z\geq1$.
\end{abstract}



\section{Introduction}\label{sec:intro}

Over the past decades, we witnessed progressive improvements in the
field of observational cosmology, for what concerns both data
acquisition and modelling. Exploiting various independent cosmological
probes, the so-called standard $\Lambda$-cold dark matter
($\mathrm{\Lambda CDM}$) model has been constrained with high levels
of accuracy and precision. Several projects have been carried on to
explore the properties of cosmic tracers at different scales, with the
primary goal of understanding the formation and evolution of the
Universe. The main properties of the large-scale structure of the
Universe have been constrained both at very high redshifts, exploiting
the Cosmic Microwave Background (CMB) power spectrum
\citep{WMAP_final_2013ApJS, Planck_Legacy_2018}, and in the local
Universe thanks to increasingly large surveys of galaxies and galaxy
clusters \citep[e.g.][]{WiggleZ_2012, 6dF_2014, Guzzo_VIPERS_2014,
  SDSS_BOSS_2017, Pacaud_2018AA}.

The unprecedented amount and quality of the data expected from the
upcoming projects will allow us to test fundamental physics, shedding 
light on questions that have remained unanswered for years. In particular, in
the era of huge galaxy survey projects, such as the Dark Energy
Survey\footnote{\url{http://www.darkenergysurvey.org}} (DES)
\citep{DES2017}, the extended Roentgen Survey with an Imaging
Telescope Array (eROSITA) satellite
mission\footnote{\url{http://www.mpe.mpg.de/eROSITA}}
\citep{Merloni2012}, the NASA Wide Field Infrared Space Telescope
(WFIRST) mission\footnote{\url{http://wfirst.gsfc.nasa.gov}}
\citep{Spergel2013}, the ESA Euclid
mission\footnote{\url{http://www.euclid-ec.org}} \citep{Laureijs_2011,
  Amendola2018}, the Large Synoptic Survey
Telescope\footnote{\url{http://www.lsst.org}} (LSST)
\citep{ivezic2008}, and the Square Kilometre Array (SKA)
\citep{Maartens_SKA_2015, Santos_SKA_2015}, we will have the
opportunity to clarify some of the main issues in the current
understanding of the Universe, such as the physical nature of dark
matter (DM) and dark energy (DE), and to test the gravity theory on
the largest scales accessible \citep[for a recent review see
  e.g.][]{Silk_2017nuco}. In fact, about $95\%$ of the content of the
Universe still remains with an unsatisfactory physical
explanation. This represents the main motivation for the forthcoming
generation of galaxy surveys, whose main goal is to achieve a better
understanding of the nature of DM and DE components. Increasingly
large and accurate maps of galaxies and other cosmic tracers will be
exploited to probe the expansion history of the Universe and the
formation of cosmic structures with unprecedented accuracy, allowing
us to robustly discriminate among alternative cosmological frameworks.

In this context, one of the most powerful tools to characterise the
spatial distribution of cosmic tracers is provided by the two-point
correlation function (2PCF), or analogously the power spectrum, which
encodes most of the information available. In particular, the
so-called redshift-space distortions (RSD) in the tracer clustering
function \citep{Jackson_FoG_1972, Kaiser_1987, Hamilton_review_1998,
  Scoccimarro_model} have been effectively exploited to test the
gravity theory on cosmological scales, providing robust constraints on
the linear growth rate of cosmic structure, using different
techniques in both configuration space \citep[e.g.][]{guzzo2000,
  reid2012, beutler2012, samushia2012, chuang2013, chuang2013b,
  delatorre2013b, samushia2014, howlett2015, okumura2016, chuang2016,
  Pezzotta_vimos_2017, mohammad2018} and Fourier space
\citep[e.g.][]{tojeiro2012, blake2012, blake2013, beutler2014}. Linear
growth rate constraints have been also obtained from the joint
analysis of galaxy clustering and weak gravitational lensing
\citep[e.g.][]{delatorre2017}, from cosmic void profiles
\citep[e.g.][]{Hamaus2016, Achitouv2017, hawken2017}, and from other
different tracers of the peculiar velocity field
\citep[e.g.][]{percival2004, davis2011, feix2015, huterer2017,
  adams2017}. Moreover, it has been shown that RSD provide a powerful
probe to constrain the mass of relic cosmological neutrinos
\citep{Marulli_2011MNRAS, Upadhye_2019JCAP} and the main parameters of
interacting DE models \citep{Marulli_2012MNRAS, Costa_2017JCAP}, as
well as help in breaking the degeneracy between modified gravity
and massive neutrino cosmologies \citep{Moresco2017, Wright_2019JCAP,
  garcia-farieta2019}.

In this paper, we present a systematic validation analysis of the main
statistical techniques currently used to constrain the linear growth
rate from redshift-space anisotropies in the 2PCF of cosmic
tracers. In \citet{Bianchi2012} and
\citet{Marulli_anisotropies_2012MNRAS, Marulli2017} we performed a
similar investigation, testing RSD likelihood modules on large mock
catalogues extracted from N-body simulations of the standard
cosmological framework. Here we extend these previous studies in many
important aspects. First, instead of modelling the two-dimensional
2PCF, we consider either the monopole and quadrupole multipole moments
of the 2PCF, or the clustering wedges, which encode most of the
information in the large-scale structure distribution.
Moreover, we investigate new RSD models based on perturbation theory,
namely the \citet{Scoccimarro_model} and \citet{TNS_model} models,
that we compare to the so-called dispersion model
\citep{Davis_1983ApJ, Peacock1996}. As in
\citet{Marulli_anisotropies_2012MNRAS}, we investigate also the impact
of redshift measurement errors, which introduce spurious small-scale
clustering anisotropies. We focus on the redshift range $0.5\lesssim
z\lesssim2$, and consider mildly non-linear scales $10<r[$\Mpch$]<55$,
where the assumptions in the RSD models considered in this work are
expected to be reliable. In addition, we investigate the impact of
considering only larger scales, $r>30$ \Mpch, where the models are
supposed to be less biased.

The paper is structured as follows. In Section \ref{sec:simulations}
we describe the set of N-body simulations employed in the analysis,
and the selected mock DM halo samples. In Section
\ref{sec:clustering}, we analyse the clustering of DM haloes in real
and redshift space, investigating the impact of redshift measurement
errors. The RSD likelihood models and the linear growth rate and bias
measurements are presented in Section \ref{sec:modelling}. Finally, in
Section \ref{sec:conclusions}, we draw our conclusions.


\section{N-body simulations and mock halo catalogues}
\label{sec:simulations}

We consider a subset of the DM halo catalogues extracted from the
publicly available \mdpl N-body simulations, which belong to the
\textsc{MultiDark} suite \citep{Riebe_2013AN_334,
  Klypin_Multidark_2016}, that is available at the \textsc{CosmoSim}
database\footnote{\url{http://www.cosmosim.org/}}. These simulations
have been widely used in recent years for different cosmological
analyses \citep[see \eg][]{vandenBosh_Multidark_2014,
  Rodirguez_Multidark_2016, Klypin_Multidark_2016,
  Vega-Ferrero_Multidark_2017, Zandanel_Multidark_2018,
  Topping_Multidark_2018, Wang_Multidark_2018,
  Ntampaka_Multidark_2019, Grannet_Multidark_2019}. The \mdpl
simulations followed the dynamical evolution of $3840^3$ DM particles,
with mass resolution of $1.51\times10^9 h^{-1}\,M_\odot$, in a
comoving box of $1000$\Mpch\, on a side, assuming a \lcdm framework
consistent with Planck constraints \citep{Planck_2014, Planck_2016}:
$\Omega_m = 0.307$, $\Omega_\Lambda = 0.693$, $\Omega_b =0.048$,
$\sigma_8 = 0.823$, $n = 0.96$ and $h = 0.678$. The DM haloes
\citep{Riebe_2013AN_334} have been identified with a
Friends-of-Friends (FoF) algorithm with a linking length of $0.2$
times the mean interparticle distance \citep{Halos_MAD_2011MNRAS}.

For the clustering analysis presented in this paper we make use of one
realisation of the halo samples per each redshift considered,
selecting only DM haloes with more than $665$ particles, which
corresponds to a minimum mass threshold of $M_{\rm min}=10^{12}
h^{-1}\,M_\odot$. The samples have been restricted in the mass range
$M_{\rm min}<M<M_{\rm max}$, where $M_{\rm max}=2\times10^{15},
1.3\times10^{15}, 7.4\times10^{14}, 5.4\times10^{14},
4.0\times10^{14}, 3.6\times10^{14}, 3.1\times10^{14}h^{-1}\,M_\odot$,
at $z=0.523,~0.740,~1.032,~1.270,~1.535,~1.771,~2.028$, respectively.
These DM haloes are more massive than the ones tipically
  hosting the faintest galaxies that will be detected in
  next-generation surveys, like e.g. Euclid and DESI. The non-linear
  RSD effects are sensitive to the tracer bias, and systematic model
  uncertainties are expected to be larger for lower bias tracers. Thus,
  the performances of the considered RSD models on forthcoming
  clustering measurements might be worse than the ones
  obtained in this paper. Higher resolution simulations and more
  realistic mock galaxy catalogues are required to test this
  hypotesis.


\section{Clustering of DM haloes}
\label{sec:clustering}
In this Section, we describe the methodology used to quantify the halo
clustering through the 2PCF, which constitutes the main subject of our
study. Specifically, we characterise the anisotropic clustering
either with the first two non-null multipole moments of the 2PCF, or
with the clustering wedges. All numerical computations in the
current Section and in the following ones have been performed with the
\textsc{CosmoBolognaLib}, a large set of {\em free software} libraries
\citep{CosmoBolognaLib}\footnote{Specifically, we used the {\small
    CosmoBolognaLib} V5.3, which includes the new implemented RSD
  likelihood modules required for the current analysis. The software
  and its documentation are freely available at the GitHub repository:
  \href{https://github.com/federicomarulli/CosmoBolognaLib}{https://github.com/federicomarulli/CosmoBolognaLib}.}.
\begin{figure*}
  \includegraphics[width=\linewidth]{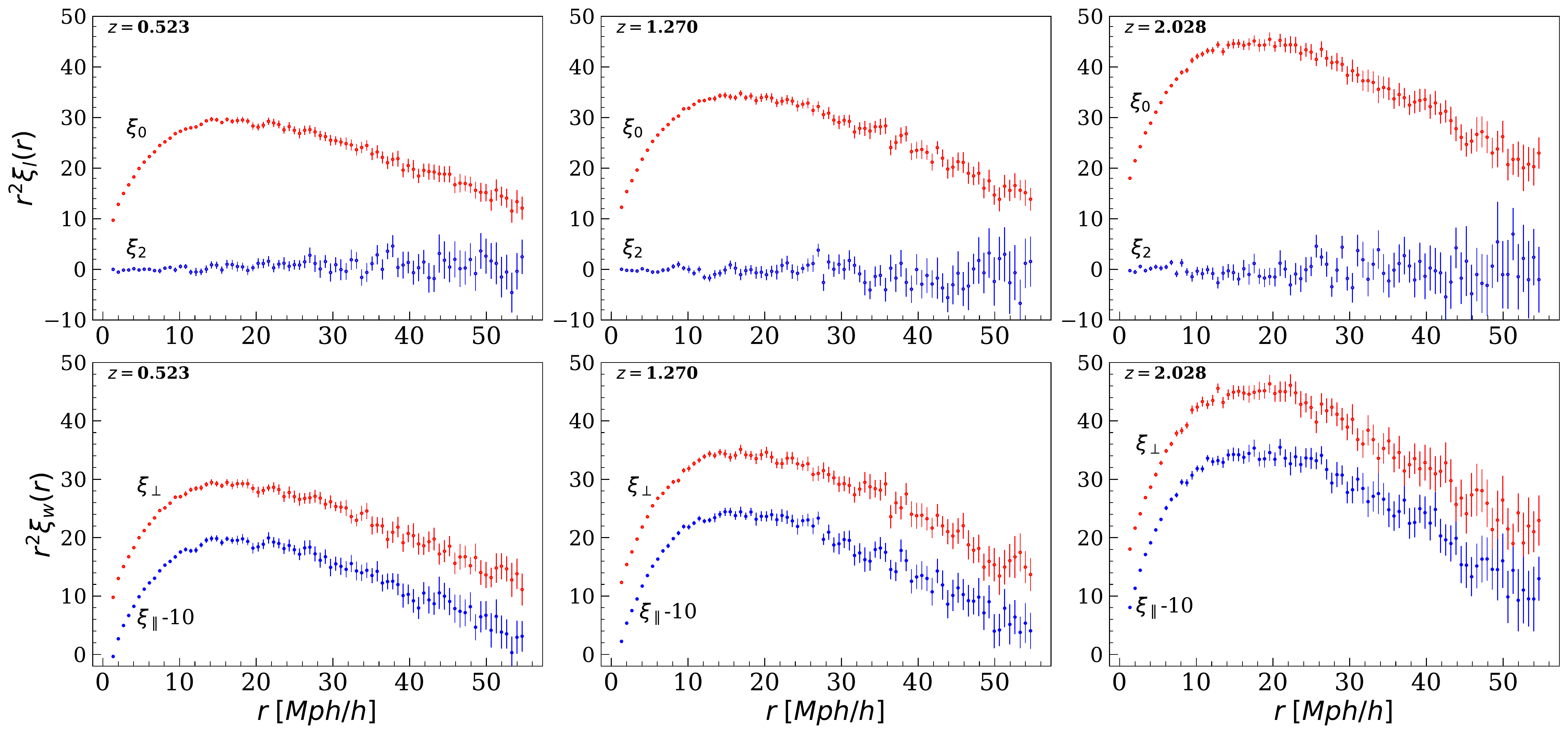}
    \caption{The real-space 2PCF of DM haloes at three different
      redshifts, as indicated by the labels. {\em Upper panels}: the
      monopole, $\xi_0$, and quadrupole, $\xi_2$, moments. {\em Bottom
        panels}: the perpendicular, $\xi_\perp$, and parallel,
      $\xi_\parallel$, wedges; the latter are shifted by $-10$, for
      clarity reasons. The error bars are computed with bootstrap
      sampling.}
    \label{fig:measure_multipoles_wedges}
\end{figure*}


\subsection{The 2PCF}
\label{subsec:2pcf}
We characterise the spatial distribution of DM haloes in the \mdpl
simulations with the 2PCF in both real space, $\xi(r,\mu)$, and
redshift space, $\xi(s,\mu)$. Specifically, we measure the full 2D
2PCF with the conventional \citet{LandySzalay_1993} estimator:
\begin{equation}
\hat{\xi}(r,\mu)=\frac{DD(r, \mu)-2DR(r, \mu) +RR(r, \mu)}{RR(r, \mu)}
\,,
\end{equation}
with $\mu$ being the cosine of the angle between the line-of-sight
(LOS) and the comoving separation $r$, and $DD(r,\mu), RR(r,\mu)$, and
$DR(r,\mu)$ being the normalised number of pairs of DM haloes in data-data,
random-random and data-random catalogues, respectively. We measure the
2PCF in the scale range from $1$ to $55$\Mpch, in $80$ linearly
spaced bins, with random samples five times larger than the halo ones,
to keep the shot noise errors due to the finite number of random
objects negligible.

The clustering anisotropies can be effectively quantified by
decomposing the full 2D 2PCF either in its multipole moments or in the
so-called wedges \citep{Kazin_2012_estimators, Sanchez2013,
  Sanchez2014, Sanchez2017}. In terms of the first non-vanishing
Legendre multipole moments, the 2D 2PCF is written as follows:
\begin{equation}
  \label{eq:ximultiexp}
  \xi(s,\mu)=\xi_0(s)L_0(\mu)+\xi_2(s)L_2(\mu)+\xi_4(s)L_4(\mu)\, ,
\end{equation}
where $L_l(\mu)$ are the Legendre polynomials of degree $l$ (i.e. $L_0
= 1$,~ $L_2=(3\mu^2-1)/2$,~$L_4 =(35\mu^4-30\mu^2+3)/8$), and the
coefficient of the expansion corresponds to the $l^{th}$ multipole
moment of the 2PCF:
\begin{equation}
  \label{eq:multipoles}
  \xi_l(s) \equiv \frac{2l +1}{2}\int_{-1}^{+1}\, \mbox{d}\mu\, \xi(s,\mu)
  L_l(\mu)\,.
\end{equation}
In this work we measure the multipole moments through the \emph{direct
  estimator}, performing the pair-counting directly in 1D bins, instead
of integrating over 2D bins as in the \emph{integrated
  estimator}. This is convenient to avoid uncertainties due to binning
effects in the numerical integration, and to optimise computational
performances. Since our random pairs do not depend on $\mu$,
\ie~$RR(r,\mu) = RR(r)$, the two estimators provide the same results
\citep[e.g.]{Kazin_2012_estimators, Marulli_XXL_2018}. In real space
the full clustering signal is encoded in the monopole moment,
$\xi_0(r)$. In redshift space the odd multipole moments vanish by
symmetry at first order. Here we focus on the first two non-null
multipole moments, that is the monopole $\xi_0(s)$ and the quadrupole
$\xi_2(s)$.

An alternative description of the clustering anisotropies is provided
by the clustering wedges, introduced by \citet{Kazin_2012_estimators},
that correspond to the angle-averaged of the \xiis over wide bins of
$\mu$:
\begin{equation}
  \label{eq:wedges}
  \xi_{w}(s) \equiv \frac{1}{\Delta\mu} \int_{\mu_{1}}^{\mu_{2}} \xi(s,
  \mu) \mathrm{d} \mu\,,
\end{equation}
where $\Delta\mu = \mu_{2}-\mu_{1}$ is the wedge width. In this work
we consider the two clustering wedges with $\Delta\mu=0.5$, that is
the transverse wedge, $\xi_\perp(s) \equiv \xi_{1/2}(\mu_{min} = 0,
s)$, and the radial (or LOS) wedge, $\xi_\parallel(s) \equiv
\xi_{1/2}(\mu_{min} = 0.5, s)$, computed in the ranges $0\leq\mu<0.5$
and $0.5\leq\mu\leq1$, respectively. The clustering wedges are related
to the multipole moments through the following equation:
\begin{equation}
  \label{eq:wedges_multipoles}
  \xi_{w}(r)=\sum_l\xi_l(s)\bar{L}_l\,,
\end{equation}
where $\bar{L}_l$ is the average value of the Legendre polynomials
over the interval $[\mu_1,\mu_2]$. Neglecting contributions
  from multipoles with $l>2$ and wedge width $\Delta\mu=0.5$,
Eq. \eqref{eq:wedges_multipoles} can be approximated as
follows \citep{Kazin_2012_estimators}:
\begin{equation}
  \label{eq:wedges_multipoles2}
  \left(\begin{array}{l}{\xi_{ \|}}
    \\ {\xi_{\perp}}\end{array}\right)=\left(\begin{array}{cc}{1} &
    {\frac{3}{8}} \\ {1} &
    {-\frac{3}{8}}\end{array}\right)\left(\begin{array}{c}{\xi_{0}}
    \\ {\xi_{2}}\end{array}\right)\,.
\end{equation}
In real space, the radial and transverse wedges are identical, and equal
to the monopole, since there are no distortions in any direction. 
\begin{figure*}
  \includegraphics[width=.45\linewidth]{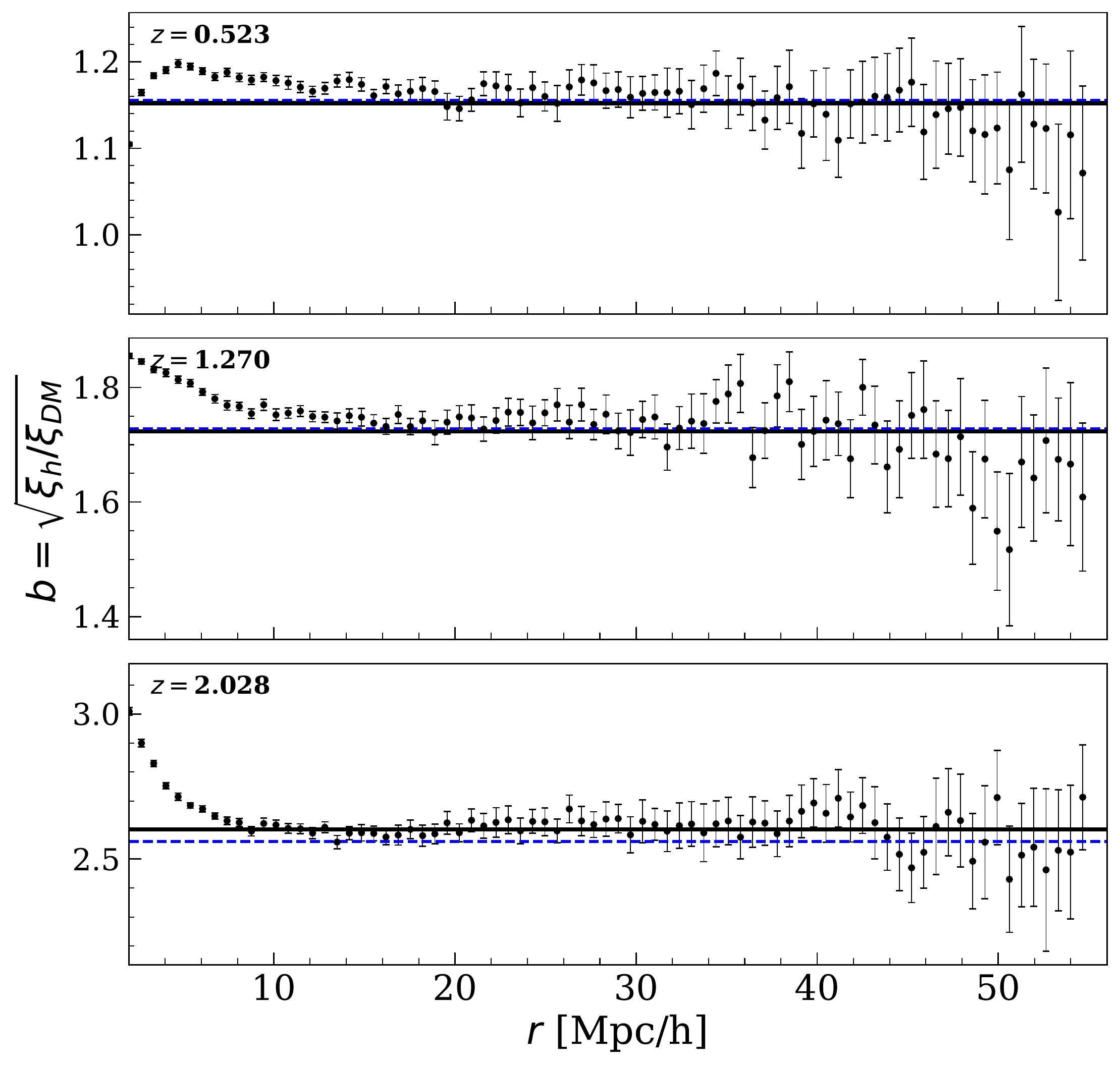}
  \includegraphics[width=.45\linewidth]{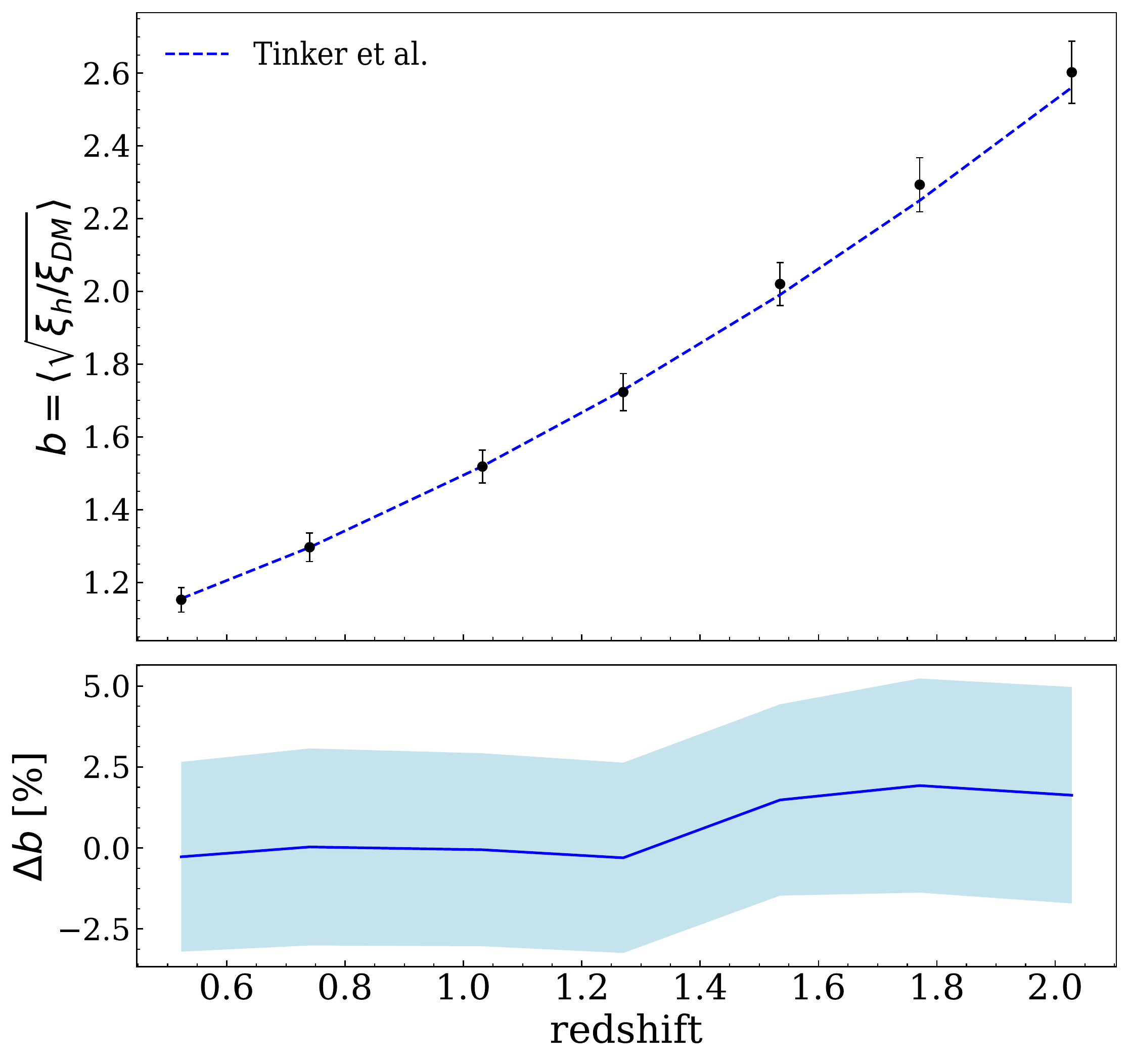}
  \caption{{\em Left panel}: the effective halo bias as a function of the
    comoving scale, at three different redshifts, as indicated by the
    labels. Dashed lines show the theoretical \lcdm effective bias
    predicted by \citet{Tinker2008, Tinker_bias_2010}, while solid 
    lines show the best-fit bias obtained from the measurements. {\em
      Right panel}: the mean effective halo bias as a function of
    redshift, computed by averaging in the scale range
    $10<r[$\Mpch$]<50$. The dashed line shows the
    \citet{Tinker2008, Tinker_bias_2010} prediction. The deviation
    between measured and theoretical effective bias values is reported
    in the bottom panel, where the shaded blue area represents the
    propagated measurement errors.}
  \label{fig:bias_evolution}
\end{figure*}

The errors on the 2PCF measurements are estimated by using the
bootstrap resampling method \citep{Efron_1979}. Firstly, the original
catalogue is divided into $27$ sub-samples, which are then re-sampled
in $100$ different data sets with replacement, then the $\xi(r,\mu)$
is measured in each one of them \citep{Barrow_1984, Ling_1986}. The
covariance matrix, $C_k(s_i, s_j)$, is computed as follows:
\begin{equation}
  C_{k}\left(s_{i}, s_{j}\right)=\frac{1}{N_{R}-1}
  \sum_{n=1}^{N_{R}}\left[\xi_{k}^{n}\left(s_{i}\right)-\overline{\xi}_{k}
    \left(s_{i}\right)\right]\left[\xi_{k}^{n}
    \left(s_{j}\right)-\overline{\xi}_{k}\left(s_{j}\right)\right] \, .
\end{equation}
The indices $i$ and $j$ run over the 2PCF bins, while $k$ refers
either to the order of the multipole moments considered, in which case
$k=l=0,~2$, or to the clustering wedges, with $k=w=0,~0.5$. In both
cases, $\bar{\xi}_k=1/N_R\sum_{n=1}^{N_R}\xi_k^n$ is the average
multipole (wedge) of the 2PCF, and $N_R=100$ is the number of
realisations obtained by resampling the catalogues with the bootstrap
method. We do not correct the inverse covariance matrix
  estimator to account for the finite number of realisations
  \citep{Hartlap2007}. This is not crucial in the context of the
  present work, which is focused on systematic errors caused by
  approximations in RSD models.


\subsection{Clustering in real space}
\label{subsec:real_space}
Figure \ref{fig:measure_multipoles_wedges} shows the real-space 2PCF
of DM haloes at three different redshifts. The upper panels show the
multipole moments, namely monopole, $\xi_0(r)$, and quadrupole,
$\xi_2(r)$. As expected, the real-space monopole moment contains the
full clustering signal, while the real-space quadrupole moment is
consistent with zero, at $1\sigma$, at all scales. The lower panels
show the perpendicular, $\xi_\perp(r)$, and parallel,
$\xi_\parallel(r)$, clustering wedges. The latter are shifted by
$-10$ for visualisation purposes. As mentioned before and as confirmed
by our results, the two wedges are statistically equal in real space,
for isotropy, and equal to the monopole moment. In all cases, the
error bars are computed with the bootstrap method.

The amplitude of the real-space clustering signal allows us to
characterise the effective halo bias, $b_{\rm eff}$, which relates the
halo clustering to the underlying mass distribution. Specifically,
$b_{\rm eff}$ can be estimated as follows:
\begin{equation}
  \label{eq:biasefflss} 
  \left\langle b_{\rm eff}(z)\right\rangle = \left\langle\sqrt{\frac{\xi_{\rm
        halo}}{\xi_{\rm DM}}}~\right\rangle \, ,
\end{equation}
where $\xi_{\rm halo}$ is the measured 2PCF of the \mdpl DM haloes,
while the DM 2PCF, $\xi_{\rm DM}$, is computed by Fourier transforming
the non-linear matter power spectrum obtained with {\small CAMB}
\citep{Lewis_2000}, which includes {\small HALOFIT}
\citep{Smith_halofit_2003, Takahashi_2012ApJ}. The bias is averaged in
the scale range $10<r[$\Mpch$]<55$.


The left panel of Fig.~\ref{fig:bias_evolution} shows the measured DM
halo bias as a function of scale, with error bars propagated from the
2PCF. The dashed blue lines correspond to the theoretical prediction
computed by averaging the linear bias, $b(M, z)$, of the selected set
of DM haloes as follows:
\begin{equation}
  \label{eq:biaseff}
  b_{\rm eff}(z) = \frac{\int_{M_{\rm min}}^{M_{\rm max}}n(M, z) b(M,
    z) d M}{\int_{M_{\rm min}}^{M_{\rm max}} n(M, z) dM},
\end{equation} 
where the mass limits $[M_{\rm min}$,~$M_{\rm max}]$ have been defined
in Section \ref{sec:simulations}, while the mass function, $n(M, z)$,
and the linear bias, $b(M, z)$, are estimated using the
\citet{Tinker2008} model and the \citet{Tinker_bias_2010} model,
respectively. The solid black lines show the best-fit bias obtained
from the measurements.

A scale-dependent behaviour of the bias can be appreciated at scales
smaller than $10$\Mpch, with deviations of about $4$\% with respect to
the theoretical linear predictions. We note that at these small scales
the assumed DM power spectrum model might not be accurate enough,
considering the measurement clustering uncertainties of this
analysis. Thus the observed scale dependence of the bias might be
partially caused by model systematics. However, this does not affect
our results, as we do not consider these scales in our analysis.  The
right panel of Fig.~\ref{fig:bias_evolution} shows the redshift
evolution of the mean effective bias, compared to the theoretical
\lcdm predictions by \citet{Tinker2008, Tinker_bias_2010}. The error
bars are computed by propagating the 2PCF errors estimated with
bootstrap resampling. Measurements appear in excellent agreement with
theoretical expectations.
\begin{figure*}
  \includegraphics[width=0.95\linewidth]{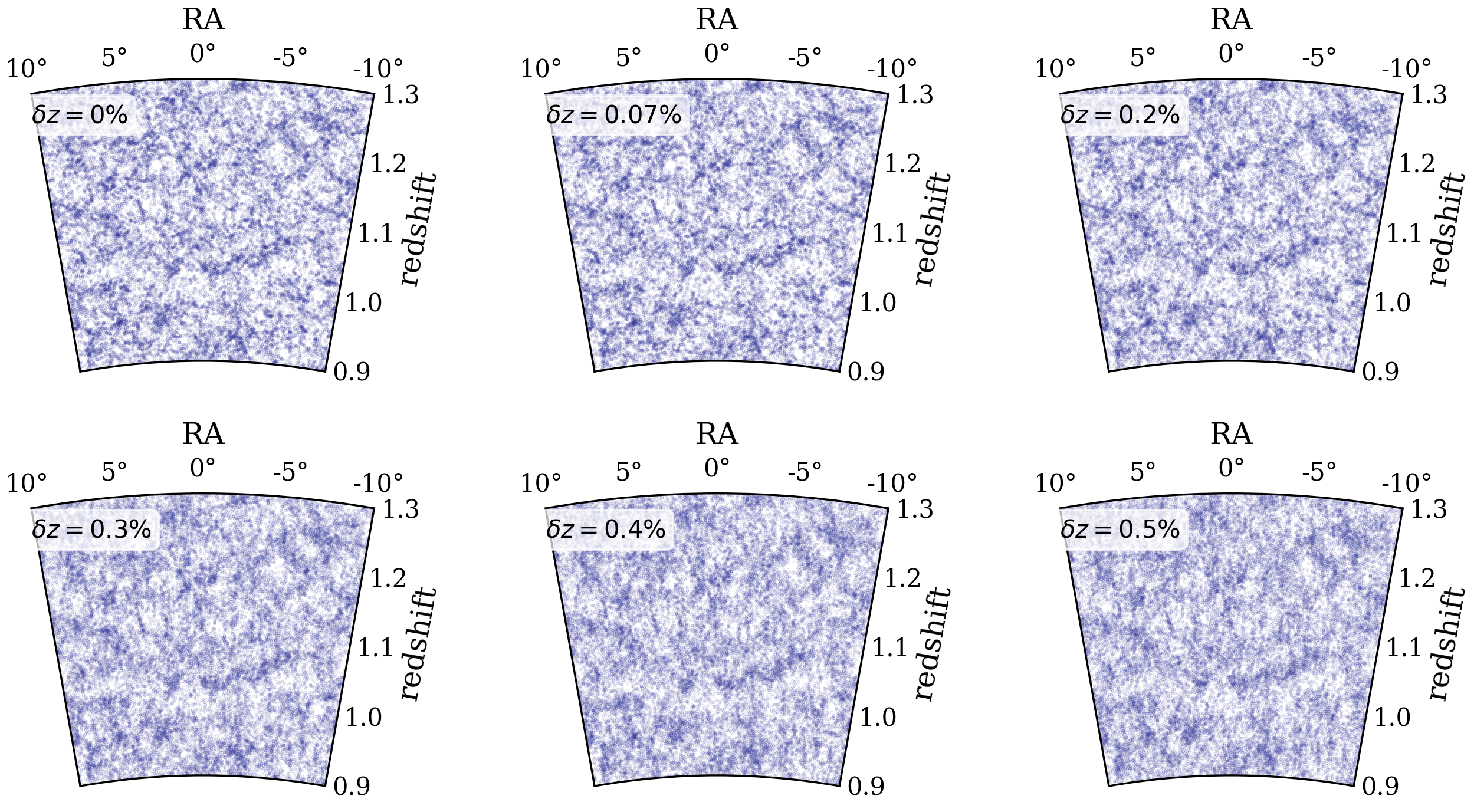}
  \caption{Spatial distribution of DM haloes in the mock sample
    corresponding to the N-body snapshot at $z=1.032$, including
    redshift measurement errors, as indicated by the labels. Only
    haloes in a $2$ degree declination slice are plotted, for
    clarity.}
  \label{fig:slice_RADECZ}
\end{figure*}


\subsection{Clustering in redshift space and dynamic distortions}
\label{subsec:redshift_space}
When comoving distances are estimated from observed redshifts, $z_{\rm
  obs}$, without correcting for the LOS peculiar velocity
contribution, the resulting clustering pattern appears
distorted. These clustering anisotropies are known as dynamic
distortions, or RSD. Specifically, $z_{\rm obs}$ can be approximated
as a combination of three terms \citep[e.g.][]{Marulli_anisotropies_2012MNRAS}: i) the cosmological redshift, $z_c$, due to the Hubble flow, ii) the change caused by the peculiar velocity along the LOS, and iii) an additional term due to the
redshift measurement errors coming from the adopted instrumentation
and calibration analysis. Neglecting the latter two terms introduces displacements between the matter distribution in real and redshift space \citep[for a review
  see][]{Hamilton_review_1998, Scoccimarro_model}.

We construct mock halo catalogues in redshift space following the same
procedure adopted by \citet{Marulli_anisotropies_2012MNRAS,
  Marulli2017}. First, we introduce a local observer at a random
position in the simulation. Then we transform the comoving coordinates
of each DM halo into polar coordinates, and estimate the observed
redshifts assuming the following relation:
\begin{equation}
  \label{eq:zobs}
  z_{\rm obs} =
  z_c+(1+z_c)\frac{\mathbf{v}\cdot\hat{\mathbf{x}}}{c}\hat{\mathbf{x}}+\frac{\sigma_v}{c}
  \,,
\end{equation}
where $\hat{\mathbf{x}}$ is a unit vector along the LOS, and $\sigma_v$
corresponds to the amplitude of a Gaussian noise in the measured
redshift expressed in $km/s$, so that the contribution of peculiar
motions is given by $\mathbf{v}_\parallel=\mathbf{v}\cdot\hat{\mathbf{x}}$. Finally,
we return back to comoving Cartesian coordinates, mimicking the
distortions in redshift space by replacing $z_c$ with $z_{\rm obs}$ to
estimate the comoving distance.  As in
\citet{Marulli_anisotropies_2012MNRAS}, we consider the following
values for the $\sigma_v$ term: ${0,~200,~500,~1000,~1250,~1500}$
km/s, which correspond to the percentage uncertainties $\delta
z=\{0,~0.07,~0.2,~0.3,~0.4,~0.5\}$\%. These values cover a sensible
range extending from the case with negligible redshift errors
($\sigma_v=0$) to the case with errors representative to those
expected from next generation spectroscopic surveys. As reference, Table
\ref{tab:redshift_errors_photometric} reports the ratios between the
$\sigma_v$ values considered in this work and the ones expected in a 
Euclid-like spectroscopic galaxy survey, that is
$\sigma_z/(1+z)\sim0.001$ \citep{Laureijs_2011}.
\begin{table}
  \centering
  \caption{Values of the Gaussian redshift errors considered in
      this work, in units of the redshift errors expected in a
      Euclid-like spectroscopic galaxy survey.}
  \begin{center}
    \begin{tabular}{|c|c|c|c|c|c|}
      \hline\hline
      & \multicolumn{5}{c}{$\sigma_v$ [km/s]} \\\cmidrule(lr){2-6}
      $z$ & 200 & 500 & 1000 & 1250 & 1500 \\
      \hline
      0.523 &   0.44 &   1.09 &   2.19 &   2.74 &   3.28 \\ 
      0.740 &   0.38 &   0.96 &   1.92 &   2.39 &   2.87 \\ 
      1.032 &   0.33 &   0.82 &   1.64 &   2.05 &   2.46 \\ 
      1.270 &   0.29 &   0.73 &   1.47 &   1.84 &   2.20 \\ 
      1.535 &   0.26 &   0.66 &   1.31 &   1.64 &   1.97 \\ 
      1.771 &   0.24 &   0.60 &   1.20 &   1.50 &   1.80 \\ 
      2.028 &   0.22 &   0.55 &   1.10 &   1.38 &   1.65 \\
      \hline
    \end{tabular}
  \end{center}
  \label{tab:redshift_errors_photometric}
\end{table}

Figure \ref{fig:slice_RADECZ} shows the spatial distribution of DM
haloes in the mock sample corresponding to the N-body snapshot at
$z=1.032$, including increasing redshift measurement errors. The slight
elongation increasing with $\sigma_v$ in the halo distribution along
the LOS due to redshift errors can be appreciated in the different
panels.

\begin{figure*}
  \includegraphics[width=0.95\linewidth]{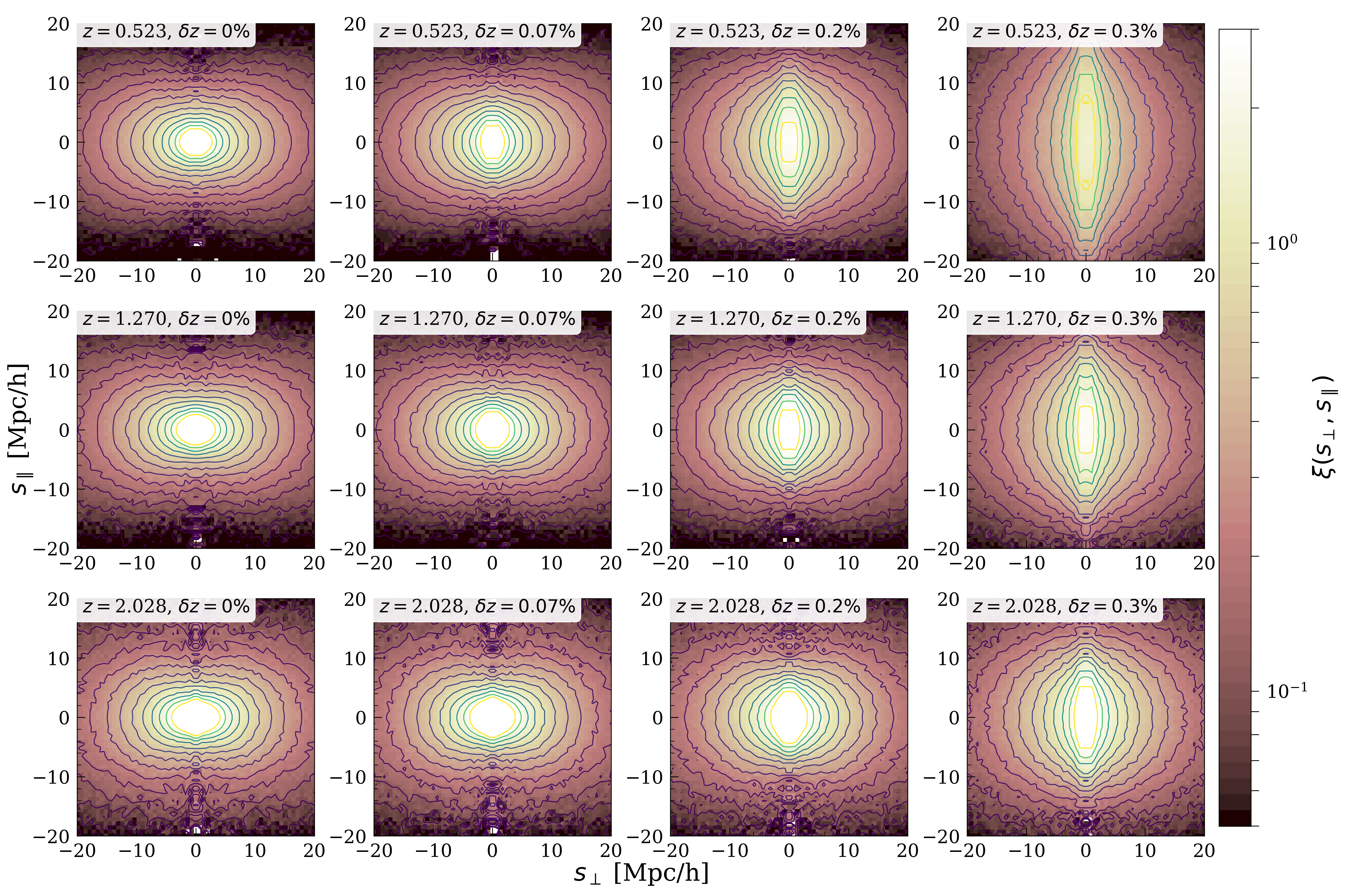}
  \caption{Iso-correlation contours of $\xi(s_\perp, s_\parallel)$, at
    the three redshifts indicated by the labels, in correspondence of
    the correlation levels \xiis$={0.05, ~0.07, ~0.09, ~0.13, ~0.18,
      ~0.24, ~0.33, ~0.45, ~0.62, ~0.85, ~1.17, ~1.6, ~2.2, ~3}$. The
    panels refer to different redshifts (rows) and different amplitudes of the redshift errors (columns), as
    indicated by the labels. The colour bar on the right side
    indicates the amplitude of \xiis.}
  \label{fig:2D_2PCF}
\end{figure*}

Figure \ref{fig:2D_2PCF} shows the 2PCF as a function of the
transverse, $s_\perp$, and parallel, $s_\parallel$, separations to the
LOS, at three different redshifts. The iso-correlation contours of
\xiis are measured in the range $[0.05,~3]$, for different values of
the redshift measurement errors, $\delta z$. As it can be seen,
redshift errors introduce spurious clustering anisotropies at small
scales, enhancing the clustering signal along the LOS, analogously to
the effect due to Fingers-of-God (FoG)
\citep{Marulli_anisotropies_2012MNRAS}.

\begin{figure*}
  \includegraphics[width=\linewidth]{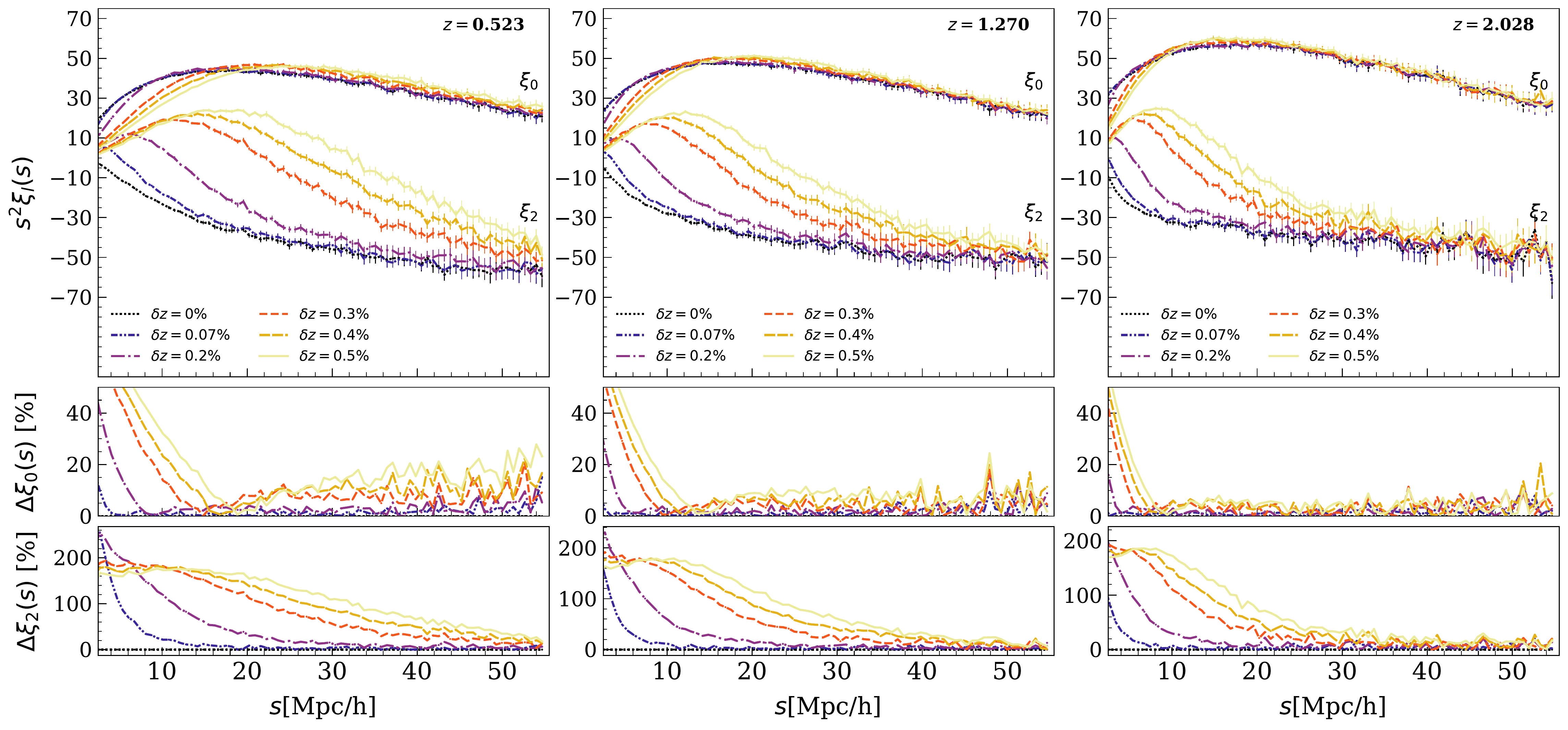}
  \caption{Redshift-space 2PCF monopole, $\xi_0$, and quadrupole,
    $\xi_2$, of the \mdpl DM haloes, at three different redshifts. The
    lines correspond to the 2PCFs measured in mock catalogues
    with different redshift errors, as indicated by the labels. The
    bottom subpanels show the relative percentage differences with
    respect to the case with no redshift errors.}
  \label{fig:measure_multipoles_ZS}
\end{figure*}

\begin{figure*}
  \includegraphics[width=\linewidth]{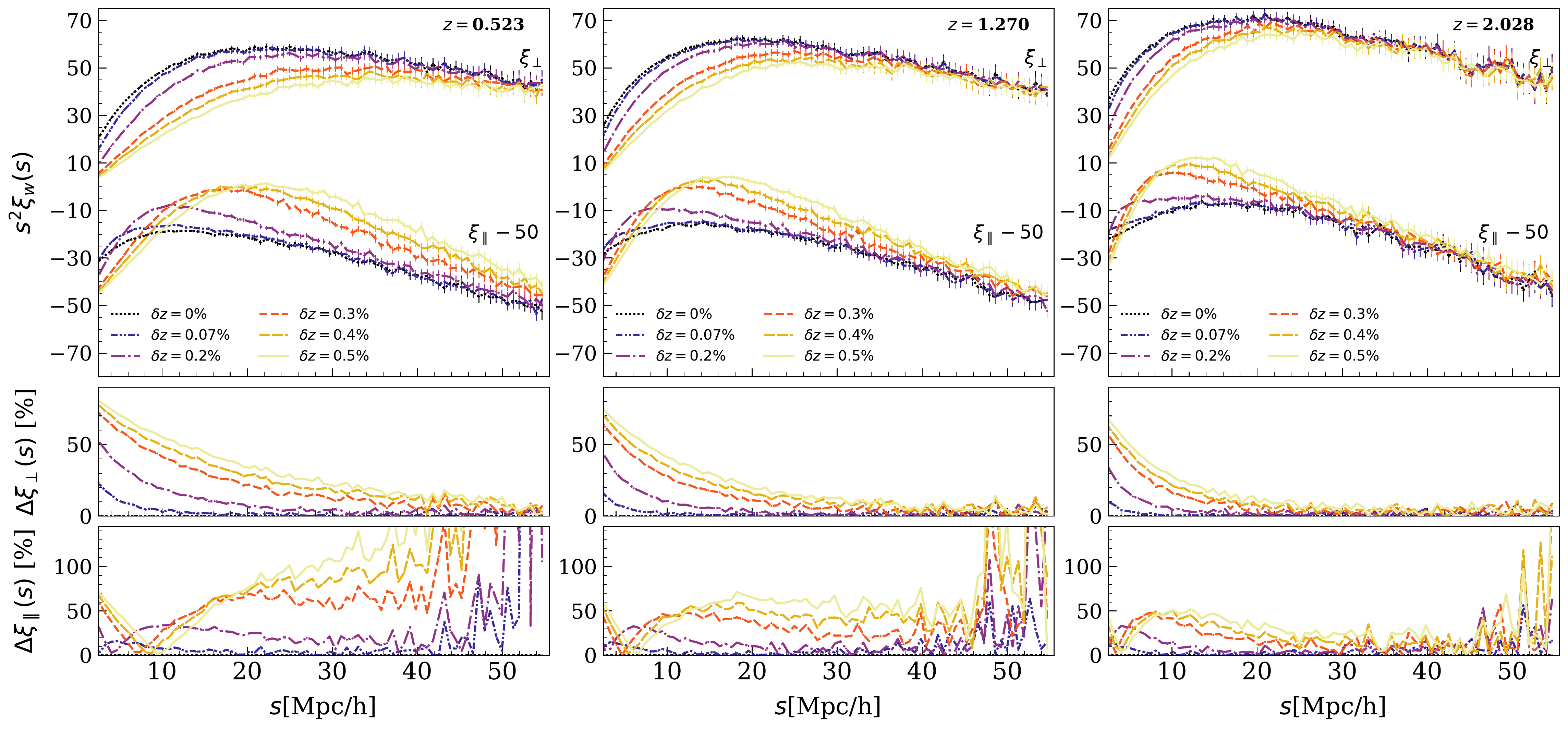}
  \caption{As Fig. \ref{fig:measure_multipoles_ZS}, but for the 
  redshift-space 2PCF perpendicular, $\xi_\perp$, and parallel, 
  $\xi_\parallel$, wedges of the \mdpl DM haloes. For clarity, 
  $\xi_\parallel$ is shifted by $-50$.}
  \label{fig:measure_wedges_ZS}
\end{figure*}

As described in Section \ref{subsec:2pcf}, it is convenient to project
the two-dimensional 2PCF, $\xi(s_\perp,s_\parallel)$, onto
one-dimensional statistics, such as the multipole moments and the
clustering wedges. In Figs. \ref{fig:measure_multipoles_ZS} and
\ref{fig:measure_wedges_ZS} we show the redshift-space monopole and
quadrupole moments, and the redshift-space radial and transverse
wedges, respectively. In agreement with
\citet{Marulli_anisotropies_2012MNRAS}, we find a progressive
  suppression of the slope of the 2PCF monopole, tending to flatness for increasingly
  larger redshift errors \citep[e.g][]{Sereno2015}. On the other hand, the quadrupole signal
increases.
The results for the clustering wedges are similar, showing a
small-scale suppression in the transverse wedge, when the redshift 
errors are included, while the radial wedge increases. As shown in
Fig. \ref{fig:2D_2PCF} and discussed in details in the next Sections,
the spurious anisotropies caused by redshift errors in the multipole
moments and wedges have a scale-dependent pattern similar to the FoG
one, caused by small-scale incoherent motions.


\begin{figure*}
  \includegraphics[width=\linewidth]{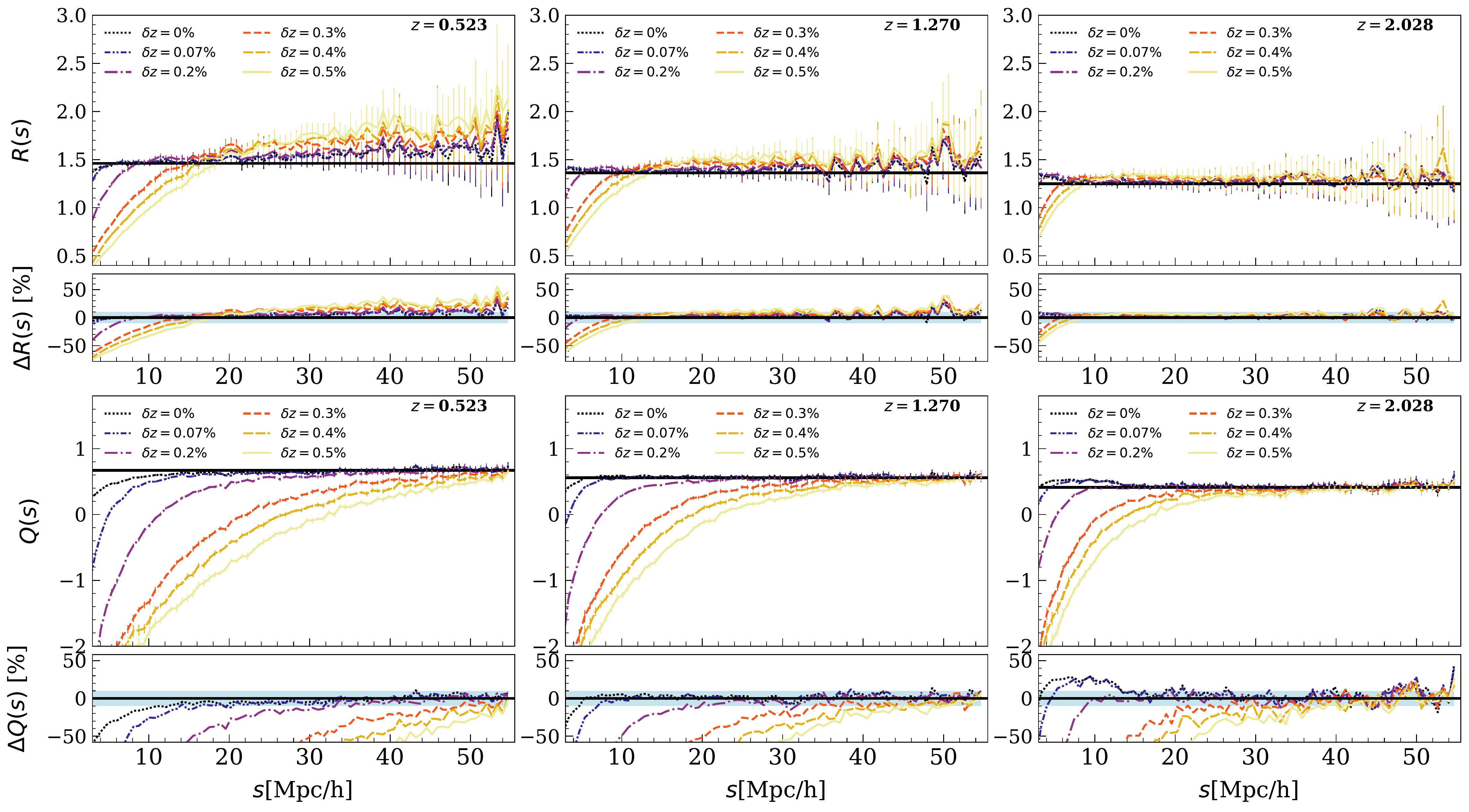}
  \caption{The ratio between the redshift-space and real-space monopole
    moments, $R(s)$ ({\em upper panels}), and between the
    redshift-space quadrupole and monopole, $Q(s)$ ({\em lower
      panels}), at three different redshifts (different columns) and for different
    redshift errors, as indicated by the labels. Horizontal lines
    represent the theoretical predictions obtained assuming the
    \citet{Tinker2008, Tinker_bias_2010} effective bias. The error bars are computed
    by propagating the 2PCF bootstrap errors and the subpanels show the 
    relative percentage differences with respect to the case with no 
    redshift errors.}
  \label{fig:multipoles_ratios}
\end{figure*}

Alternative statistics that can be used to quantify the impact of
redshift errors in the clustering pattern are the ratio between the
redshift-space and real-space monopole, $R(s)$, and the ratio between
the redshift-space quadrupole and monopole, $Q(s)$. In the linear
regime, these quantities can be written as follows:
\begin{eqnarray}
  \label{eq:xi_ratios_multipoles}
  R(s)&=&\frac{\xi_0(s)}{\xi_0(r)} = 1+ \frac{2\beta}{3} +
  \frac{\beta^2}{5},
  \\ Q(s)&=&\frac{\xi_2(s)}{\xi_0(s)-\frac{3}{s^3}\int^s_0ds'\xi(s')s'^2}
  = \frac{\frac43\beta+\frac47\beta^2}{1+ \frac{2\beta}{3} +
    \frac{\beta^2}{5}}\,,
\end{eqnarray}
where $\xi_0$ and $\xi_2$ are the redshift-space monopole and
quadrupole of the 2PCF, respectively, and $\beta$ is the linear
distortion parameter defined as $\beta\equiv f(z)/b(z)$, with $f(z)$ being the linear growth rate. Figure
\ref{fig:multipoles_ratios} shows the measured $R(s)$ and $Q(s)$
statistics, as a function of redshift errors, compared to the
theoretical predictions derived by assuming the \citet{Tinker2008,
  Tinker_bias_2010} effective bias. As it can be seen, we find a good
agreement between measurements and theoretical predictions in the case
without redshift errors, for both estimators, at large enough scales
(beyond $\sim10$\Mpch). Redshift errors introduce scale-dependent
distortions in both these statistics. In particular, their effect is
to increase (decrease) the $R(s)$ ratio above (below) a characteristic
scale, whereas the $Q(s)$ is reduced, especially at small scales.


\section{Modelling redshift-space distortions}
\label{sec:modelling}
In this Section, we describe the models used to parameterise the RSD
in the 2PCF multipoles and wedges. Then we derive constraints on
$f\sigma_8$ and $b\sigma_8$ parameters for each mock catalogue
constructed from the \mdpl simulations, investigating the effect of
possible redshift errors. The multipole moments are modelled as
follows:
\begin{equation}
  \label{eq:mult_model}
  \xi_l(s) = i^l\int_{-\infty}^{\infty}\frac{\mbox{d}
    k}{2\pi^2}k^2P_l(k)j_l(ks)\,,
\end{equation}
where $j_l$ are the spherical Bessel functions, and $P_l(k)$ are the
power spectrum multipoles:
\begin{equation}
  \label{eq:Pkmult_model}
  P_l(k) = \frac{2l+1}{2}\int_{-1}^1\mbox{d} \mu\, P^s(k,
  \mu)L_l(\mu)\,.
\end{equation}
We consider three widely-used RSD models to estimate the
redshift-space 2D power spectrum $P^s(k, \mu)$:

\begin{itemize}
  \item Dispersion model \citep{Peacock1996}:
    \begin{equation}
      \label{eq:dispersion_model}
      P^s(k, \mu) = D(k, f, \mu, \sigma_{12})\, \left(1+\frac{f}{b} \mu^2
      \right)^2\,b^2\,P_{\delta\delta}(k)\,,
    \end{equation}
    where the second term on the right-hand side of the equation
    describes the distortions caused by the large-scale coherent
    peculiar motions \citep{Kaiser_1987}, $P_{\delta\delta}(k)$ is the
    matter power spectrum, and $D(k, f, \mu, \sigma_{12})$ is a damping
    factor that characterises the incoherent peculiar motions at small
    scales. In this work, we consider both the Gaussian and the  
    Lorentzian forms of the damping factor, as already done in previous works \citep[see
      e.g.][]{Scoccimarro_model, TNS_model, Marulli_anisotropies_2012MNRAS, Xu_2012A, Xu_2012B,
      Zheng_2017JCAP}:
    \begin{equation}
	    \label{eq:streamming}
		D(k, f, \mu, \sigma_{12}) = 
		\begin{cases}
			\exp \left[-k^2f^2\mu^2\sigma_{12}^2\right],   & \text{ Gaussian,} \\
			\frac{1}{(1+k^2f^2\mu^2 \sigma_{12}^2)}, & \text{Lorentzian.}
		\end{cases}
	\end{equation}   
  \item Scoccimarro model \citep{Scoccimarro_model}: this model
    considers the density and velocity divergence fields separately to
    account for their non-linear mode coupling:
    \begin{multline}
      P^{s}(k, \mu) = D(k, f, \mu, \sigma_{12})\left(b^{2} P_{\delta
        \delta}(k)+2 f b \mu^{2} P_{\delta \theta}(k)+ \right.\\ \left.+f^{2} \mu^{4} P_{\theta \theta}(k)\right)\,,
    \end{multline}
    where $P_{\delta\theta}$ and $P_{\theta\theta}$ are the
    density-velocity divergence cross-spectrum and the velocity
    divergence auto-spectrum, respectively. In the linear regime,
    both $P_{\delta\theta}$ and $P_{\theta\theta}$ tend to
    $P_{\delta\delta}$.
    

\item TNS model \citep{TNS_model}: besides taking into account the
  non-linear mode coupling between the density and velocity divergence
  fields, this model introduces also additional terms to correct for
  systematics at small scales:
  \begin{multline}
    \label{eq:TNS_model}
    P^{s}(k, \mu) = D(k, f, \mu, \sigma_{12})\biggl(b^{2} P_{\delta
      \delta}(k)+2 f b \mu^{2} P_{\delta \theta}(k)+ \\ +f^{2} \mu^{4}
    P_{\theta \theta}(k)+C_{A}(k, \mu, f, b)+C_{B}(k,\mu, f, b)\biggr)\,.
  \end{multline}
  Following \citet{TNS_model} and \citet{delaTorre_2012MNRAS}, we
  express the correction terms of the TNS model derived from the Standard
  Perturbation Theory (SPT), $C_A$ and $C_B$, in terms of the basic
  statistics of density $\delta$ and velocity divergence
  $\theta(\mathbf{k}) \equiv[-i \mathbf{k} \cdot
    \mathbf{v}(\mathbf{k})] /[a f(a) H(a)]$. Specifically, they can be
  written as follows:
  \begin{align}
    \begin{split}
      \label{eq:TNS_A_term}
      C_A(k, \mu) &=(k \mu f) \int \frac{d^{3} \boldsymbol{p}}{(2
        \pi)^{3}} \frac{p_{z}}{p^{2}}\\ &
      \times\left[B_{\sigma}(\boldsymbol{p},\boldsymbol{k}-\boldsymbol{p},-\boldsymbol{k})-B_{\sigma}(\boldsymbol{p},
        \boldsymbol{k},-\boldsymbol{k}-\boldsymbol{p})\right]\,,
    \end{split}\\
    \label{eq:TNS_B_term}
    C_B(k, \mu) &=(k \mu f)^{2} \int \frac{d^{3} \boldsymbol{p}}{(2
      \pi)^{3}} F(\boldsymbol{p}) 
      F(\boldsymbol{k}-\boldsymbol{p})\,,\\    
    \shortintertext{with}
    \label{eq:TNS_integral_term}
    F(\boldsymbol{p})&=\frac{p_{z}}{p^{2}}\left[P_{\delta \theta}(p)+f
      \frac{p_{z}^{2}}{p^{2}} P_{\theta \theta}(p)\right]\,,
  \end{align}
  and $B_\sigma$ being the cross-bispectrum.
  The $C_A$ and $C_B$ terms are proportional to $b^3$ and $b^4$,
  respectively, and can be re-written as a power series expansion of
  $b$, $f$ and $\mu$, and their respective contributions to the total
  power spectrum. For a detailed explanation on the perturbation
  theory calculations of these correction terms, see Appendix A of
  \citet{TNS_model}, while for what concerns the correlation function
  and the dependence of the spatial bias of the considered tracers, see
  Appendix A of \citet{delaTorre_2012MNRAS}.
  
\end{itemize}
  
The $P_{\delta\delta}$, $P_{\delta\theta}$ and $P_{\theta\theta}$
terms can be computed directly from perturbation theory (Eulerian,
Lagrangian or Time renormalisation) or, alternatively, using fitting
formulae \citep[see e.g.][]{Jennings_2012, Pezzotta_vimos_2017,
  Bel_2019}. In this paper we adopt the former approach, estimating
the terms of the total power spectrum using the SPT, which consists of
expanding the statistics of interest as a sum of infinite terms, each
one corresponding to a $n$-loop correction \citep[see
  e.g.][]{Gil_2012JCAP}. In particular, we consider corrections up to
$1$-loop order, thus the power spectrum can be written as
follows:
\begin{equation}
  \label{eq:SPT_1loop}
  P^{\mathrm{SPT}}(k) = P^{(0)}(k)+P^{(1)}(k)=P^{(0)}(k)+2
  P_{13}(k)+P_{22}(k)\,,
\end{equation}
where the 0-loop correction term, $P^{(0)}(k)$, corresponds to the
linear power spectrum and the one-loop contribution, $P^{(1)}(k)$,
consists of the sum of two terms, $P_{13}(k)$ and
$P_{22}(k)$. To estimate the power spectrum at $z>0$, we
  rescale Eq.~\eqref{eq:SPT_1loop} by the linear growth factor,
  $D(z)$, as follows: $P_{xy}(k,z) = [D(z)/D(0)]^2P_{11}(k) +
  [D(z)/D(0)]^4 [P_{22}(k) + 2P_{13}(k)]$, where ${x,y}$ is either
  $\delta$ or $\theta$ \citep[for details on these terms see
  e.g.][]{Bernardeau_2002PhR, Gil_2012JCAP}. We compute the quantities
in Eq. \eqref{eq:SPT_1loop} with the \texttt{CPT
  Library} \footnote{\url{http://www2.yukawa.kyoto-u.ac.jp/~atsushi.taruya/cpt_pack.html}}.

We exploit a full Markov Chain Monte Carlo (MCMC) statistical analysis
to estimate posterior distribution constraints on the three free RSD
model parameters $\left[f\sigma_8,\, b\sigma_8,\, \sigma_{12}\right]$. We consider a
standard Gaussian likelihood, defined as follows:
\begin{equation}
  -2 \ln \mathcal{L}=\sum_{i, j=1}^{N}\left[\xi_{k}^{D}(s_{i}) -
    \xi_{k}^{M}(s_{i})\right] C_{k}^{-1}(s_{i},
  s_{j})\left[\xi_{k}^{D}(s_{j}) - \xi_{k}^{M}(s_{j})\right]\,,
\end{equation}
with $N$ being the number of bins at which the multipole moments and
the wedges are computed, and the superscripts $D$ and $M$ referring to
data and model, respectively.

\begin{figure*}  
  \includegraphics[width=0.85\linewidth]{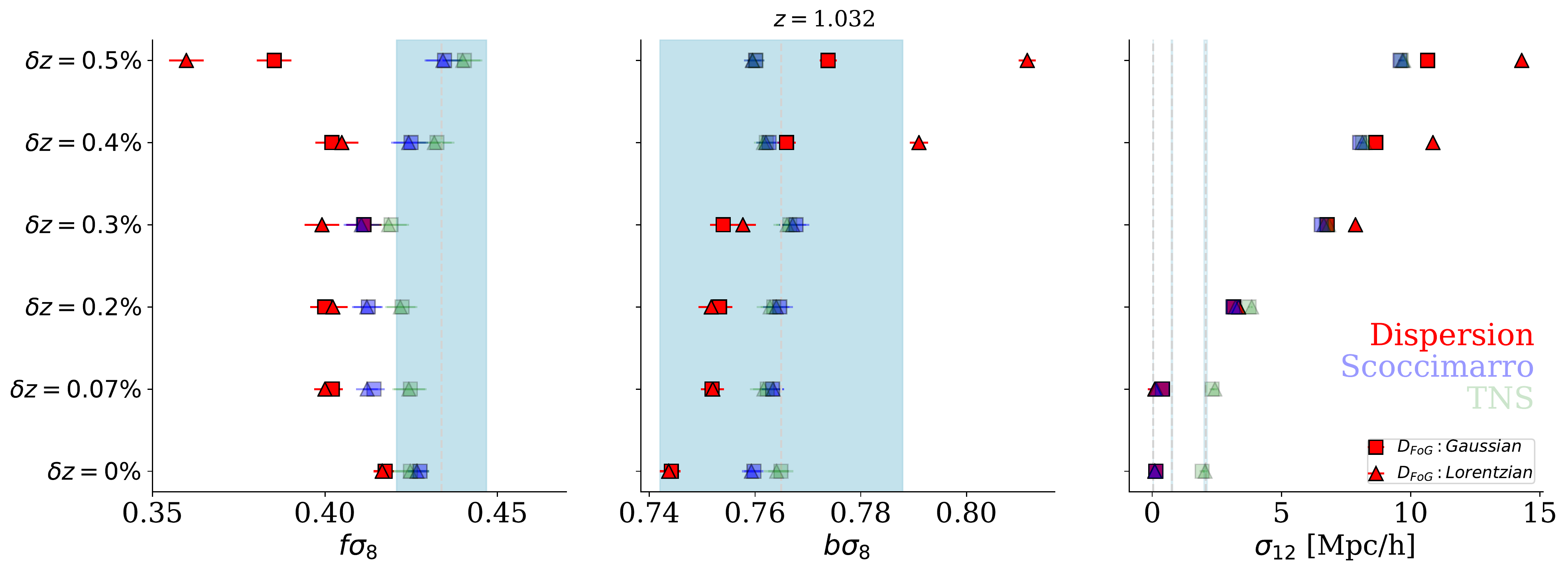}\\
  \includegraphics[width=0.85\linewidth]{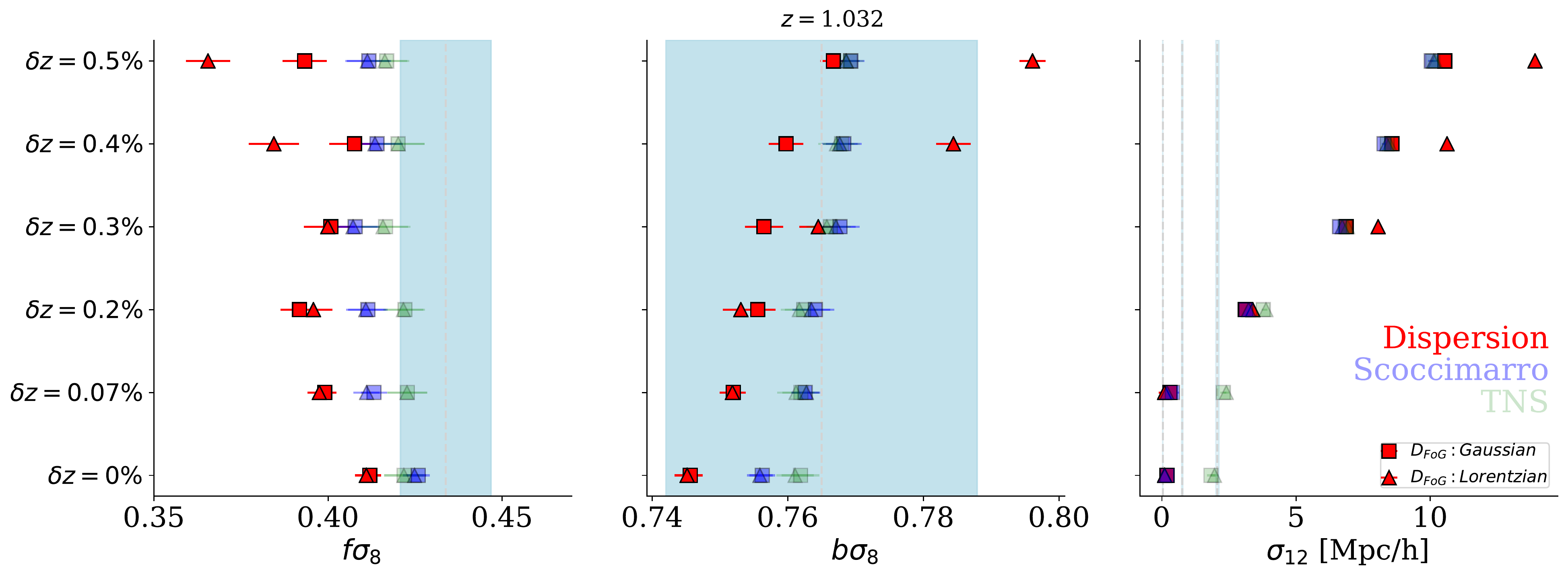}
  \caption{Best-fit constraints on [$f\sigma_8$, $b\sigma_8$,
      $\sigma_{12}$] of the \mdpl mock catalogue at $z=1.032$, assuming
    the Gaussian and Lorentzian form of the damping factor for
    different values of redshift errors. The dispersion, Scoccimarro
    and TNS models are differentiated by colour, as labelled, and the
    error bars show the $68\%$ marginalised posterior
    uncertainties. The vertical lines show the theoretical
    predictions -- the linear growth rate is computed as
    $f=\Omega_M(z)^\gamma$; the bias, $b$, is computed by assuming the
    \citet{Tinker2008, Tinker_bias_2010} effective bias model; the
    prediction for $\sigma_{12}$ is obtained from the MCMC analysis
    with only $\sigma_{12}$ as free parameter, while all the other
    parameters are fixed at their theoretical values. {\em Upper
      panel}: results from the redshift-space monopole and quadrupole
    moments; {\em lower panel}: results from the perpendicular and
    parallel wedges.}
  \label{fig:Plot_fores_Lorentz_Gauss}
\end{figure*}

We perform the MCMC analysis on all the \mdpl mock halo catalogues to
get the global evolution of the constrained parameters. First we
compare the constraints on $f\sigma_8$, $b\sigma_8$ and $\sigma_{12}$
at $z=1.032$, obtained with the Gaussian and Lorentzian damping
factors. The results are shown in
Fig. \ref{fig:Plot_fores_Lorentz_Gauss} for the redshift-space
multipole moments and clustering wedges. As it can be appreciated, the
systematic errors are lower when the damping factor is modelled with a
Gaussian function, as expected since redshifts errors are modelled as
Gaussian variables and their effects are captured by the
  damping term in the models, in agreement with
  \citet{Marulli_anisotropies_2012MNRAS}. Thus, in the following we
will adopt the Gaussian form. The fluctuations in the
  results are not statistically significant, but deserve further
  investigations with higher resolution simulations.

\begin{figure*}
  \includegraphics[width=0.95\linewidth]{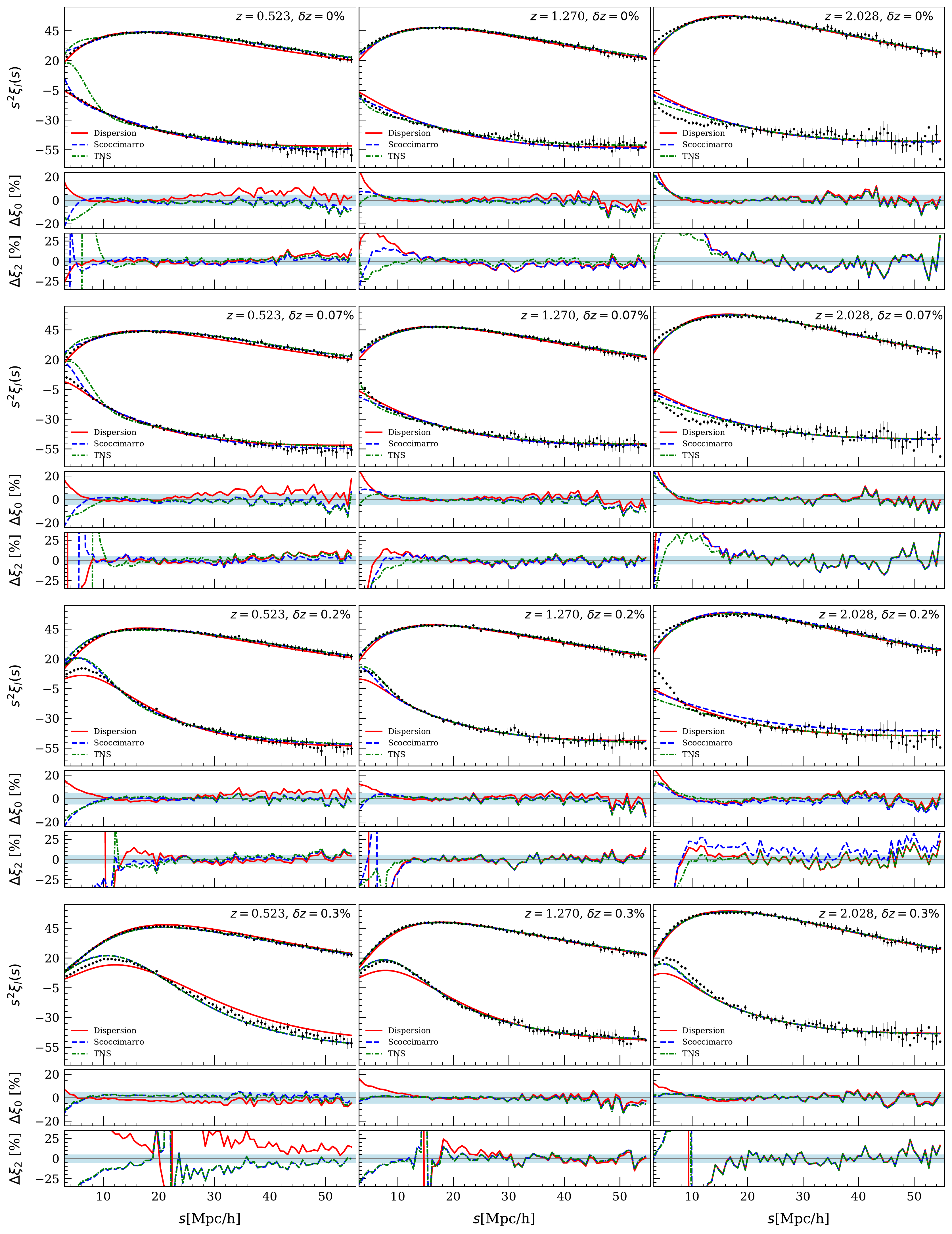}
  \caption{Redshift-space monopole, $\xi_0$, and quadrupole, $\xi_2$,
    moments of the \mdpl mock catalogues, compared to the best-fit
    models -- dispersion model (red solid), Scoccimarro model 
    (dashed blue) and TNS
    model (dash-dotted green). The results are shown at three different redshifts (different columns), and for different measurement redshift errors (different rows), as indicated by the labels. The subpanels show the relative percentage differences with
    respect to the measurements.}
  \label{fig:modelling_multipoles_bestfit}
\end{figure*}

\begin{figure*}
  \includegraphics[width=0.95\linewidth]{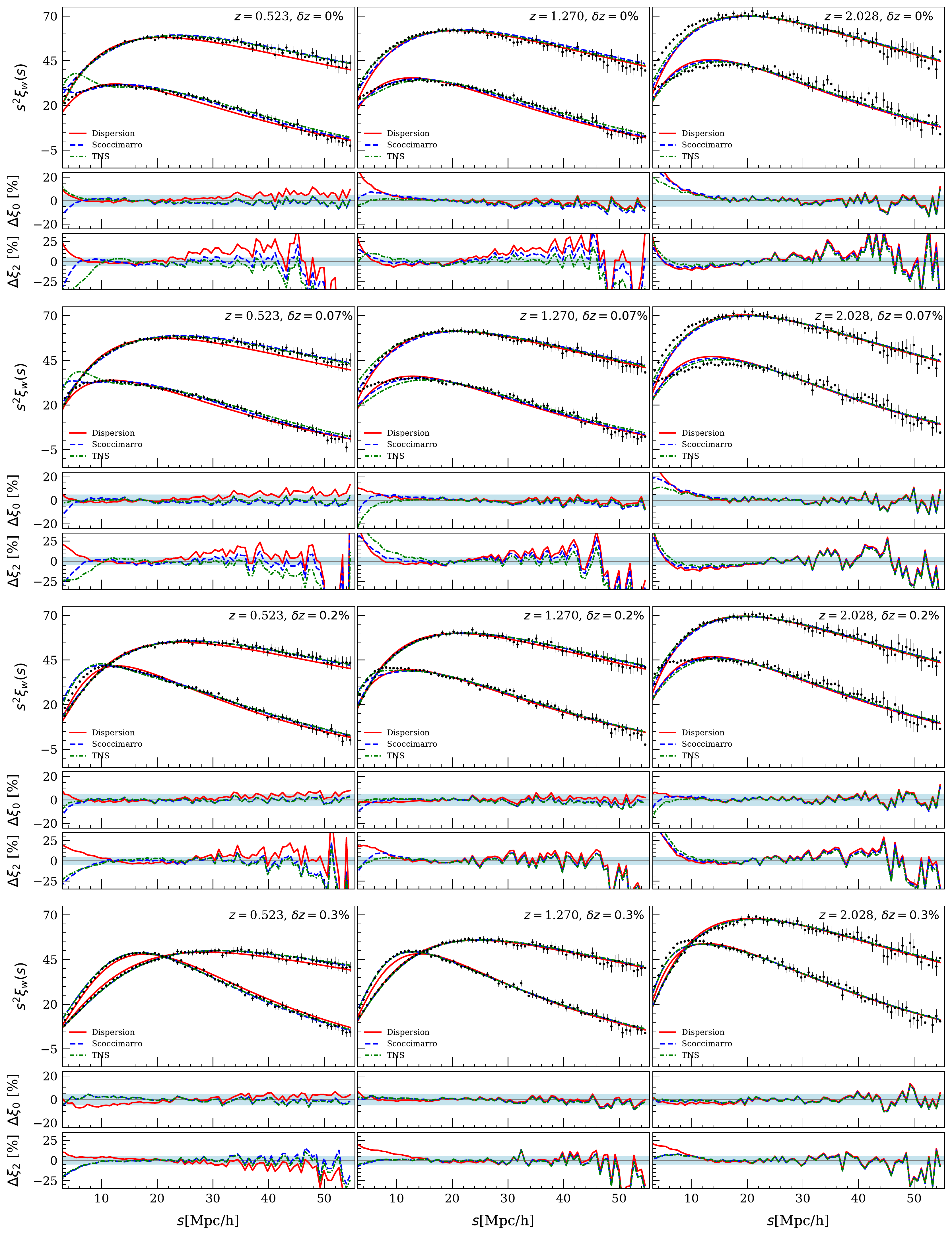}
  \caption{As Fig.~\ref{fig:modelling_multipoles_bestfit} but for the 
  redshift-space perpendicular, $\xi_\perp$, and parallel, 
  $\xi_\parallel$, wedges of the \mdpl mock catalogues.}
  \label{fig:modelling_wedges_bestfit}
\end{figure*}

Figures \ref{fig:modelling_multipoles_bestfit} and
\ref{fig:modelling_wedges_bestfit} show the measured multipole moments
and the clustering wedges compared to best-fit model predictions for
the dispersion, Scoccimarro and TNS models, at $z=0.523, ~1.032,
~2.028$, and for different redshift measurement errors. We find good
agreement between the best-fit models and the measured statistics on
scales down to about 10\Mpch, for both multipole moments and
clustering wedges, also when we include redshift errors in the
measurements. Overall, the dispersion model is the one that deviates
the most at small scales, especially when multipole moments are
considered, whereas the two SPT-based models considered in this work
fit the data better, in both statistics. In particular, at scales
larger than 10\Mpch, the percentage differences between the TNS model
and the measurements are lower than about $3\%$ and $5\%$ for the
monopole and the quadrupole, respectively. While they are lower than
about $3\%$ and $7\%$ for the perpendicular wedge and the parallel
wedge, respectively.

\begin{figure*}
  \includegraphics[width=\linewidth]{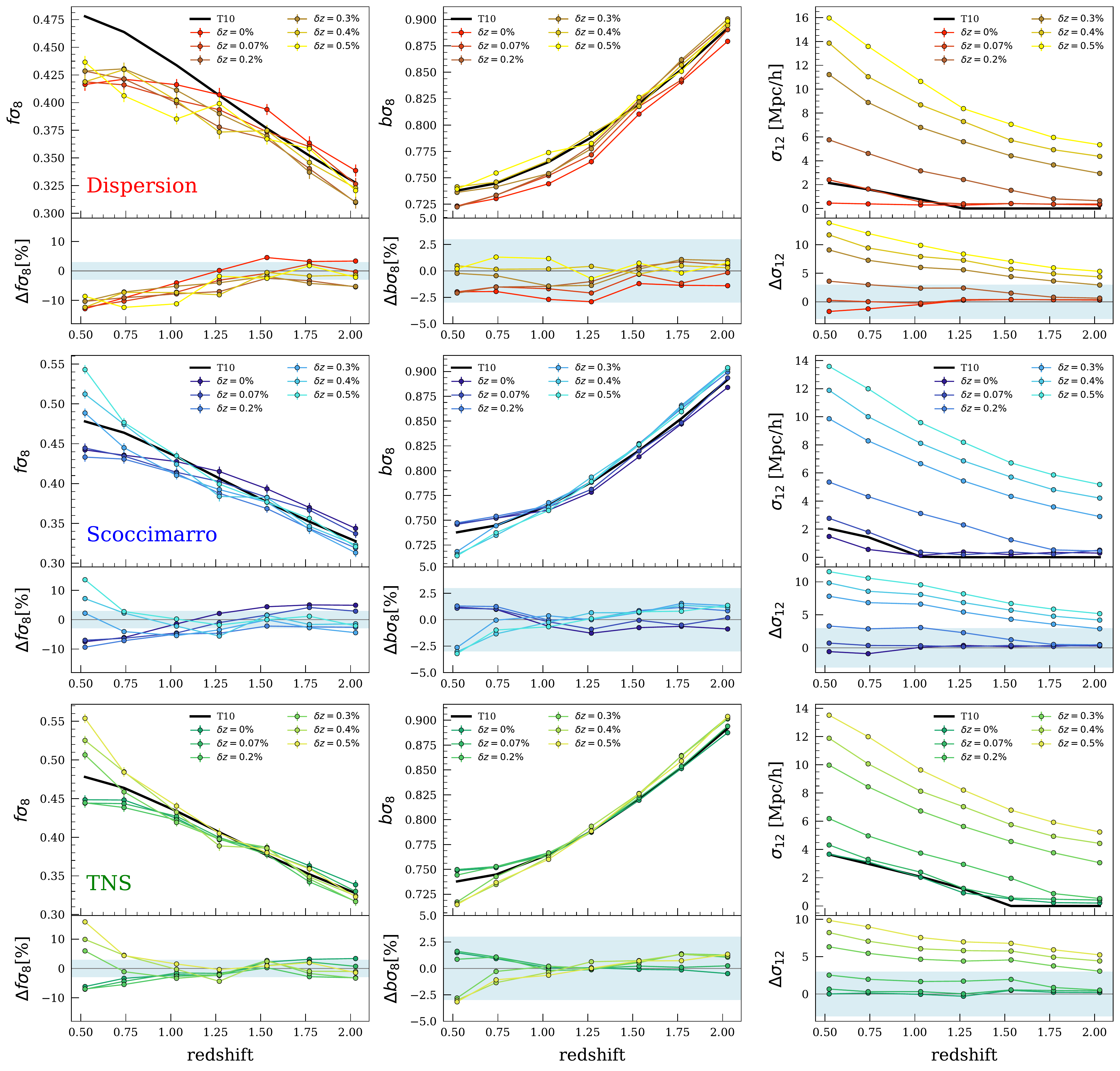}
  \caption{Best-fit constraints on [$f\sigma_8$, $b\sigma_8$,
      $\sigma_{12}$] obtained from the redshift-space monopole and
    quadrupole moments, as a function of redshift (different columns),
    and for different values of redshift errors, as indicated by the
    labels. The error bars show the $68\%$ marginalised posterior
    uncertainties. The black solid lines show the theoretical predictions --
    the linear growth rate is computed as $f=\Omega_M(z)^\gamma$; the
    bias, $b$, is computed by assuming the \citet{Tinker2008,
      Tinker_bias_2010} effective bias model; the prediction for
    $\sigma_{12}$ is obtained from the MCMC analysis with only
    $\sigma_{12}$ as free parameter, while all the other parameters
    are fixed at their theoretical values.  {\em Upper panels}:
    dispersion model; {\em central panels}: Scocimarro model; {\em
      lower panel}: TNS model. The subpanels show the relative
    percentage differences with respect to the theoretical
    prediction.}
  \label{fig:evolution_parameters_mcmc_with_distortions_in_redshift_multipol}
\end{figure*}

\begin{figure*}
  \includegraphics[width=\linewidth]{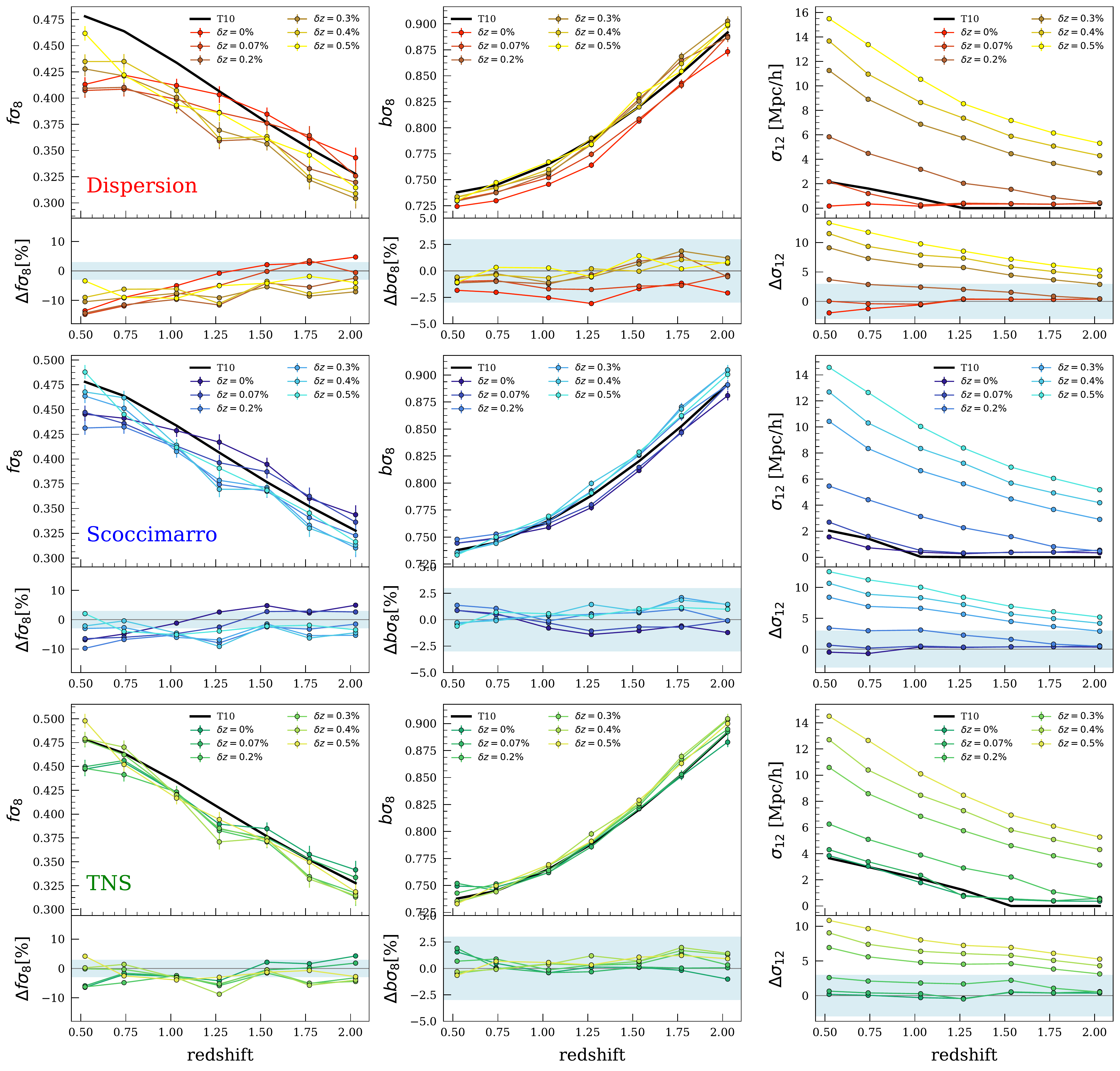}
  \caption{The same as
    Fig.~\ref{fig:evolution_parameters_mcmc_with_distortions_in_redshift_multipol},
    but using perpendicular and parallel clustering
    wedges. }
  \label{fig:evolution_parameters_mcmc_with_distortions_in_redshift_wedges}
\end{figure*}

The marginalised posterior constraints on the parameters
$\left[f\sigma_8,\, b\sigma_8,\, \sigma_{12}\right]$, as a function of
redshift, are reported in
Figs.~\ref{fig:evolution_parameters_mcmc_with_distortions_in_redshift_multipol}
and
\ref{fig:evolution_parameters_mcmc_with_distortions_in_redshift_wedges},
for multipole moments and clustering wedges, respectively. The solid
black lines represent the theoretical predictions. In particular,
$b\sigma_8$ is computed assuming the \citet{Tinker2008,
  Tinker_bias_2010} effective bias model, while the pairwise velocity
dispersion, $\sigma_{12}$, corresponds to the best-fit value obtained
when the remaining parameters are fixed to the theoretical
expectations.

In the case with no redshift errors, we find a systematic bias in the
$f\sigma_8$ constraints of about $10\%$ at low redshifts, $z<1$, for
the dispersion model, in agreement with previous works
\citep[e.g.][]{Bianchi2012, Marulli_anisotropies_2012MNRAS,
  Marulli2017}. The Scoccimarro and TNS model provide more accurate
constraints, with a systematic bias of about $8\%$ and $5\%$,
respectively.  At high redshifts, $z\geq1$, the agreement between
$f\sigma_8$ measurements and the expected values improves. In
particular, the Scoccimarro model recovers $f\sigma_8$ within $4\%$,
while the TNS model within $3\%$. The constraints on $b\sigma_8$
are overall in good agreement for all models, being the TNS model the
one with the lowest deviation with respect to the theoretical
expectations, which is found to be less than $2\%$ at all redshifts
considered.

As we have seen in Fig. \ref{fig:2D_2PCF}, the spurious anisotropies
caused by Gaussian redshift errors are similar to the FoG
distortions. The combined effects of redshift errors and FoG are thus
parameterised by the single damping term of the RSD models. Indeed, as
shown in
Figs.~\ref{fig:evolution_parameters_mcmc_with_distortions_in_redshift_multipol}
and
\ref{fig:evolution_parameters_mcmc_with_distortions_in_redshift_wedges},
the estimated value of the $\sigma_{12}$ parameter of the damping term
systematically increases as redshift errors increase. At $z\geq 1$,
the $f\sigma_8$ and $b\sigma_8$ constraints are not significantly
affected by the introduction of Gaussian redshift errors, up to
$\delta z = 0.5 \%$. On the other hand, at lower redshifts the impact
is more significant, at all redshift errors considered.

\begin{figure*}
  \includegraphics[width=\linewidth]{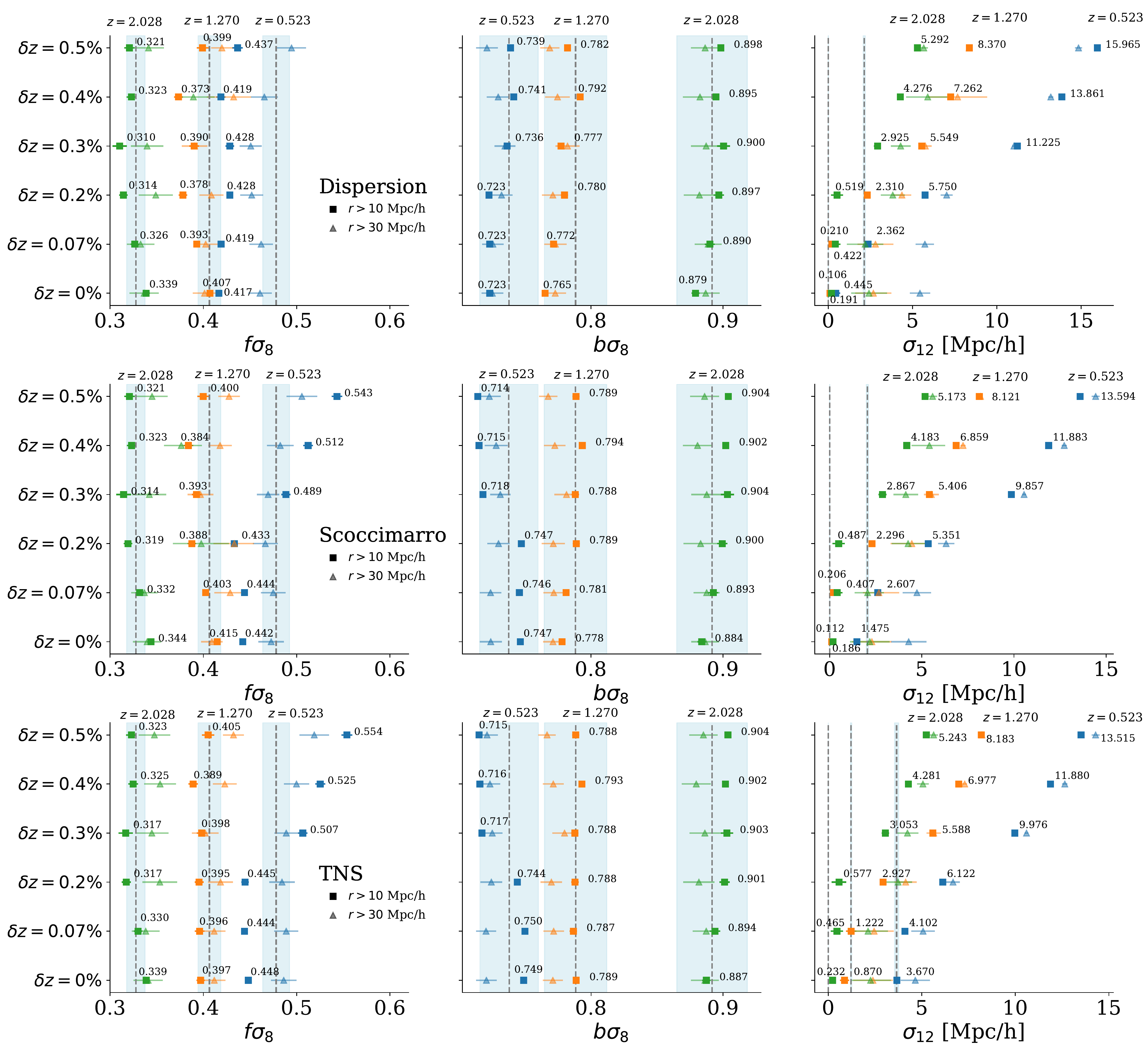}
  \caption{Marginalised posterior constraints at $68\%$ confidence
    level for $f\sigma_8$ (first column), $b\sigma_8$ (central column)
    and $\sigma_{12}$ (last column), obtained from the MCMC analysis of
    the redshift-space monopole and quadrupole moments. The results
    are shown at three different redshifts as labelled $z=0.523$, $z=1.270$
    and $z=2.028$, for the dispersion model (first
    row), the Scoccimarro model (central row), and the TNS model
    (bottom row), as labelled. The vertical black lines are centred on
    theoretical expectations, with the shaded area reporting the $3\%$
    region, for comparison.}
  \label{fig:constraints_multipoles}
\end{figure*}

\begin{figure*}
  \includegraphics[width=\linewidth]{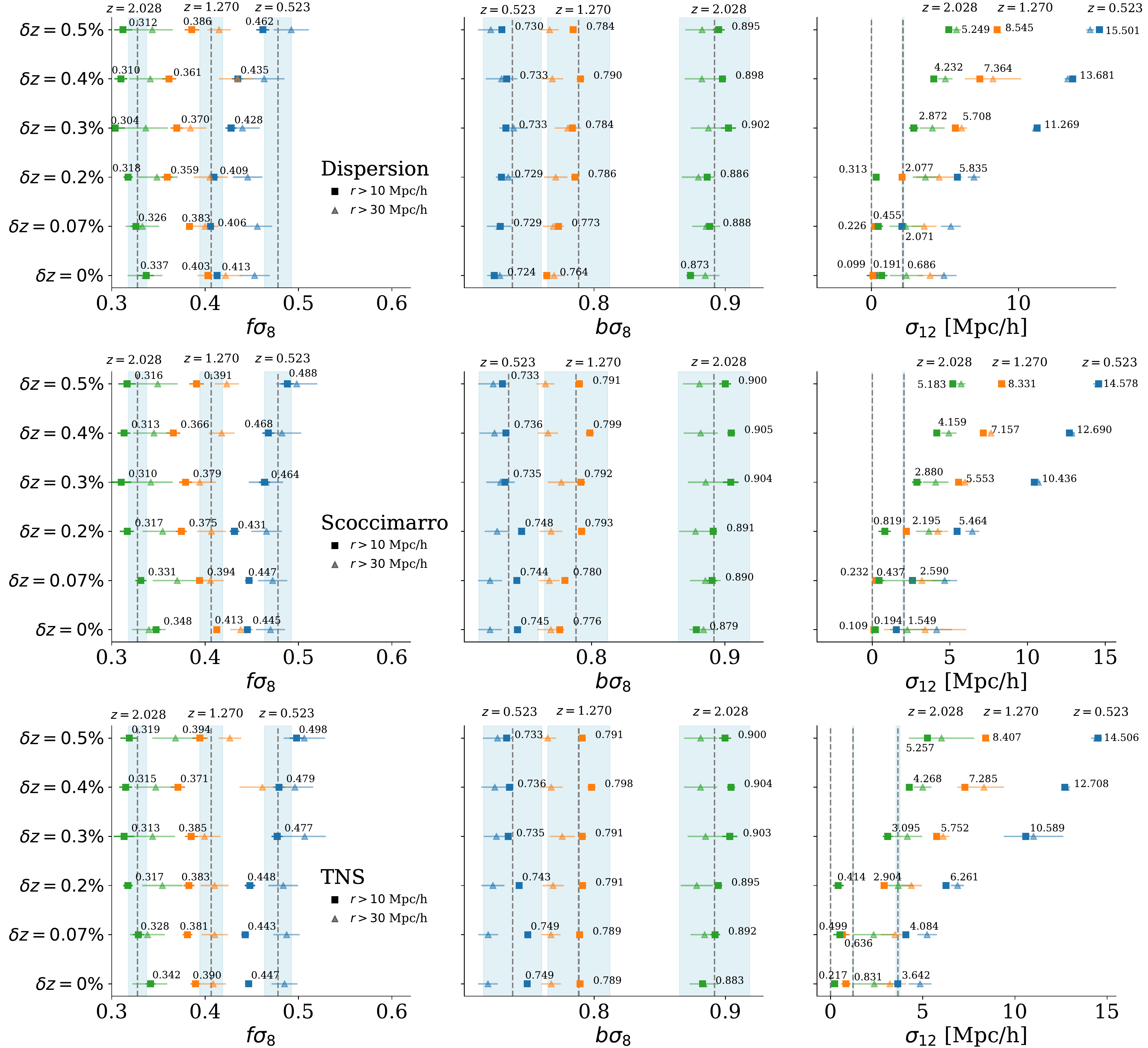}
  \caption{As Fig.~\ref{fig:constraints_multipoles}, but
    using the redshift-space perpendicular, $\xi_\perp$, and parallel, 
    $\xi_\parallel$, clustering wedges.}
  \label{fig:constraints_wedges}
\end{figure*}

Figures \ref{fig:constraints_multipoles} and
\ref{fig:constraints_wedges} summarise our main results, showing the
marginalised posterior constraints at $68\%$ confidence level for
$f\sigma_8$, $b\sigma_8$ and $\sigma_{12}$, obtained from the MCMC
analysis of the redshift-space monopole and quadrupole moments, and of
the perpendicular and parallel clustering wedges,
respectively. Moreover, Figs. \ref{fig:constraints_multipoles} and
\ref{fig:constraints_wedges} compare the results obtained by fitting
the 2PCF statistics in the comoving scale range $10<r[$\Mpch$]<55$ to
the ones obtained at scales $r>30$ \Mpch. As expected, while the
statistical uncertainties are larger in the latter scale, the
systematic discrepancies are slightly reduced. In particular, the
discrepancies of the TNS model on both the growth rate and the linear
bias are reduced below $3\%$, at $z<1.5$, for redshift errors up to
$\delta z \sim 0.3\%$. On the other hand, at larger redshifts it seems
more convenient to consider in the analysis also the small scales,
which can be reliably described by all the RSD models considered.


\section{Conclusions}
\label{sec:conclusions}

We presented a systematic analysis of state-of-the-art statistical
methods to infer cosmological constraints on the linear growth rate
from RSD in the 2PCF. This work follows from the analyses presented in
\citet{Bianchi2012} and \citet{Marulli_anisotropies_2012MNRAS,
  Marulli2017}. The two main improvements of the current study with
respect to the latter are that i) we considered both the monopole and
quadrupole moments of the 2PCF, as well as the perpendicular and
parallel clustering wedges, and ii) we compared three RSD models, that
is the dispersion model, the Scoccimarro model and the TNS model,
investigating the impact of Gaussian redshift errors on the linear
growth rate and bias constraints. The analysis has been performed in
the redshift range $0.5\lesssim z\lesssim2$, and in the comoving scale
range $10<r[$\Mpch$]<55$.

The main results of this analysis can be summarised as follows:
\begin{itemize}
\item At $z<1$, the linear growth rate measured with the dispersion
  model is underestimated by about $10\%$, in agreement with previous
  findings; the Scoccimarro and TNS models provide slightly better
  constraints, with a systematic bias of about $8\%$ and $5\%$,
  respectively.
\item As expected, limiting the analysis at $r>30$ \Mpch, the
  statistical uncertainties become larger, while the systematic
  discrepancies are slightly reduced. In particular, the systematics
  of the TNS model on both the growth rate and the linear bias are
  reduced below $3\%$, at $z<1.5$, for redshift errors up to $\delta z
  \sim 0.3\%$.
\item At $z\geq1$, all the RSD models considered provide constraints
  in good agreement with expectations. The TNS model is the one which
  performs better, with growth rate uncertainties below about $3\%$.
\item Gaussian redshift errors introduce spurious anisotropies, whose
  effect combines with the one of the small-scale incoherent motions
  responsible of the FoG distortions. This effect is captured by the
  damping factor of the RSD model considered, and can be marginalised over 
  in the statistical analysis, not introducing statistically
  significant bias in the RSD constraints, especially at $z\geq1$.
\end{itemize}

Overall, we find that the TNS model is the best one among the RSD
models considered, in agreement with previous analyses
\citep[e.g.][]{Pezzotta_vimos_2017}. The linear growth rate can be
recovered within about $3\%$ of accuracy in the redshift range
$1<z<2$, typical of next generation galaxy survey missions, like
Euclid \citep{Laureijs_2011}, even in the presence of Gaussian
redshift errors up to $\delta z=0.5 \%$, which are greater than those
expected from forthcoming spectroscopic galaxy surveys (see Table
\ref{tab:redshift_errors_photometric}). Though this accuracy is good
enough for clustering analyses in current redshift surveys, the RSD
models have to be further improved not to introduce significant
systematics in RSD constraints from next generation galaxy surveys,
which aim at mapping the cosmic structure growth rate with statistical
uncertainties below few percent. Moreover, the mass
  resolution of the N-body simulations considered in this work is too
  low to model the DM haloes typically hosting the faintest galaxies
  that will be detected in next-generation surveys. As the systematic
  model uncertainties are expected to be larger for lower bias
  tracers, the performances of the considered RSD models on
  forthcoming clustering measurements could be even worse than the
  ones reported here. Higher resolution simulations are required to
  investigate this issue and provide more realistic forecasts on
  systematic uncertainties.

\section*{Acknowledgements}
We acknowledge the grants ASI n.I/023/12/0, ASI-INAF n. 2018-23-HH.0 
and PRIN MIUR 2015 ``Cosmology and Fundamental Physics: illuminating 
the Dark Universe with Euclid".  The CosmoSim database used in this 
paper is a service by the Leibniz-Institute for Astrophysics Potsdam 
(AIP). The MultiDark database was developed in cooperation with the 
Spanish MultiDark Consolider Project CSD2009-00064. We
would also like to thank the referee for helping to improve and clarify the paper. JEGF acknowledges
financial support from ``Convocatoria Doctorados Nacionales 757 de 
COLCIENCIAS''. 

\bibliographystyle{mnras}
\bibliography{modellingPT.bib}

\begin{thebibliography}{}
\makeatletter
\relax
\def\mn@urlcharsother{\let\do\@makeother \do\$\do\&\do\#\do\^\do\_\do\%\do\~}
\def\mn@doi{\begingroup\mn@urlcharsother \@ifnextchar [ {\mn@doi@}
  {\mn@doi@[]}}
\def\mn@doi@[#1]#2{\def\@tempa{#1}\ifx\@tempa\@empty \href
  {http://dx.doi.org/#2} {doi:#2}\else \href {http://dx.doi.org/#2} {#1}\fi
  \endgroup}
\def\mn@eprint#1#2{\mn@eprint@#1:#2::\@nil}
\def\mn@eprint@arXiv#1{\href {http://arxiv.org/abs/#1} {{\tt arXiv:#1}}}
\def\mn@eprint@dblp#1{\href {http://dblp.uni-trier.de/rec/bibtex/#1.xml}
  {dblp:#1}}
\def\mn@eprint@#1:#2:#3:#4\@nil{\def\@tempa {#1}\def\@tempb {#2}\def\@tempc
  {#3}\ifx \@tempc \@empty \let \@tempc \@tempb \let \@tempb \@tempa \fi \ifx
  \@tempb \@empty \def\@tempb {arXiv}\fi \@ifundefined
  {mn@eprint@\@tempb}{\@tempb:\@tempc}{\expandafter \expandafter \csname
  mn@eprint@\@tempb\endcsname \expandafter{\@tempc}}}

\bibitem[\protect\citeauthoryear{{Achitouv}, {Blake}, {Carter}, {Koda}  \&
  {Beutler}}{{Achitouv} et~al.}{2017}]{Achitouv2017}
{Achitouv} I.,  {Blake} C.,  {Carter} P.,  {Koda} J.,   {Beutler} F.,  2017,
  \mn@doi [\prd] {10.1103/PhysRevD.95.083502}, \href
  {https://ui.adsabs.harvard.edu/abs/2017PhRvD..95h3502A} {95, 083502}

\bibitem[\protect\citeauthoryear{{Adams} \& {Blake}}{{Adams} \&
  {Blake}}{2017}]{adams2017}
{Adams} C.,  {Blake} C.,  2017, \mn@doi [\mnras] {10.1093/mnras/stx1529}, \href
  {https://ui.adsabs.harvard.edu/abs/2017MNRAS.471..839A} {471, 839}

\bibitem[\protect\citeauthoryear{{Alam} et~al.,}{{Alam}
  et~al.}{2017}]{SDSS_BOSS_2017}
{Alam} S.,  et~al., 2017, \mn@doi [\mnras] {10.1093/mnras/stx721}, \href
  {https://ui.adsabs.harvard.edu/abs/2017MNRAS.470.2617A} {470, 2617}

\bibitem[\protect\citeauthoryear{{Amendola} et~al.,}{{Amendola}
  et~al.}{2018}]{Amendola2018}
{Amendola} L.,  et~al., 2018, \mn@doi [Living Reviews in Relativity]
  {10.1007/s41114-017-0010-3}, \href
  {http://adsabs.harvard.edu/abs/2018LRR....21....2A} {21, 2}

\bibitem[\protect\citeauthoryear{{Barrow}, {Bhavsar}  \& {Sonoda}}{{Barrow}
  et~al.}{1984}]{Barrow_1984}
{Barrow} J.~D.,  {Bhavsar} S.~P.,   {Sonoda} D.~H.,  1984, \mn@doi [\mnras]
  {10.1093/mnras/210.1.19P}, \href
  {http://adsabs.harvard.edu/abs/1984MNRAS.210P..19B} {210, 19P}

\bibitem[\protect\citeauthoryear{{Bel}, {Pezzotta}, {Carbone}, {Sefusatti}  \&
  {Guzzo}}{{Bel} et~al.}{2019}]{Bel_2019}
{Bel} J.,  {Pezzotta} A.,  {Carbone} C.,  {Sefusatti} E.,   {Guzzo} L.,  2019,
  \mn@doi [\aap] {10.1051/0004-6361/201834513}, \href
  {https://ui.adsabs.harvard.edu/abs/2019A&A...622A.109B} {622, A109}

\bibitem[\protect\citeauthoryear{{Bennett} et~al.,}{{Bennett}
  et~al.}{2013}]{WMAP_final_2013ApJS}
{Bennett} C.~L.,  et~al., 2013, \mn@doi [\apjs] {10.1088/0067-0049/208/2/20},
  \href {https://ui.adsabs.harvard.edu/abs/2013ApJS..208...20B} {208, 20}

\bibitem[\protect\citeauthoryear{{Bernardeau}, {Colombi}, {Gazta{\~n}aga}  \&
  {Scoccimarro}}{{Bernardeau} et~al.}{2002}]{Bernardeau_2002PhR}
{Bernardeau} F.,  {Colombi} S.,  {Gazta{\~n}aga} E.,   {Scoccimarro} R.,  2002,
  \mn@doi [\physrep] {10.1016/S0370-1573(02)00135-7}, \href
  {https://ui.adsabs.harvard.edu/abs/2002PhR...367....1B} {367, 1}

\bibitem[\protect\citeauthoryear{{Beutler} et~al.,}{{Beutler}
  et~al.}{2012}]{beutler2012}
{Beutler} F.,  et~al., 2012, \mn@doi [\mnras]
  {10.1111/j.1365-2966.2012.21136.x}, \href
  {https://ui.adsabs.harvard.edu/abs/2012MNRAS.423.3430B} {423, 3430}

\bibitem[\protect\citeauthoryear{{Beutler} et~al.,}{{Beutler}
  et~al.}{2014}]{beutler2014}
{Beutler} F.,  et~al., 2014, \mn@doi [\mnras] {10.1093/mnras/stu1051}, \href
  {https://ui.adsabs.harvard.edu/abs/2014MNRAS.443.1065B} {443, 1065}

\bibitem[\protect\citeauthoryear{{Bianchi}, {Guzzo}, {Branchini}, {Majerotto},
  {de la Torre}, {Marulli}, {Moscardini}  \& {Angulo}}{{Bianchi}
  et~al.}{2012}]{Bianchi2012}
{Bianchi} D.,  {Guzzo} L.,  {Branchini} E.,  {Majerotto} E.,  {de la Torre} S.,
   {Marulli} F.,  {Moscardini} L.,   {Angulo} R.~E.,  2012, \mn@doi [\mnras]
  {10.1111/j.1365-2966.2012.22110.x}, \href
  {http://adsabs.harvard.edu/abs/2012MNRAS.427.2420B} {427, 2420}

\bibitem[\protect\citeauthoryear{{Blake} et~al.}{{Blake}
  et~al.}{2012}]{blake2012}
{Blake} C.,  et~al., 2012, \mn@doi [\mnras] {10.1111/j.1365-2966.2012.21473.x},
  \href {http://adsabs.harvard.edu/abs/2012MNRAS.425..405B} {425, 405}

\bibitem[\protect\citeauthoryear{{Blake} et~al.}{{Blake}
  et~al.}{2013}]{blake2013}
{Blake} C.,  et~al., 2013, \mn@doi [\mnras] {10.1093/mnras/stt1791}, \href
  {http://adsabs.harvard.edu/abs/2013MNRAS.436.3089B} {436, 3089}

\bibitem[\protect\citeauthoryear{{Campbell} et~al.,}{{Campbell}
  et~al.}{2014}]{6dF_2014}
{Campbell} L.~A.,  et~al., 2014, \mn@doi [\mnras] {10.1093/mnras/stu1198},
  \href {https://ui.adsabs.harvard.edu/abs/2014MNRAS.443.1231C} {443, 1231}

\bibitem[\protect\citeauthoryear{{Chuang} \& {Wang}}{{Chuang} \&
  {Wang}}{2013}]{chuang2013}
{Chuang} C.-H.,  {Wang} Y.,  2013, \mn@doi [\mnras] {10.1093/mnras/stt1290},
  \href {http://adsabs.harvard.edu/abs/2013MNRAS.435..255C} {435, 255}

\bibitem[\protect\citeauthoryear{{Chuang} et~al.,}{{Chuang}
  et~al.}{2013}]{chuang2013b}
{Chuang} C.-H.,  et~al., 2013, \mn@doi [\mnras] {10.1093/mnras/stt988}, \href
  {http://adsabs.harvard.edu/abs/2013MNRAS.433.3559C} {433, 3559}

\bibitem[\protect\citeauthoryear{{Chuang} et~al.,}{{Chuang}
  et~al.}{2016}]{chuang2016}
{Chuang} C.-H.,  et~al., 2016, \mn@doi [\mnras] {10.1093/mnras/stw1535}, \href
  {https://ui.adsabs.harvard.edu/abs/2016MNRAS.461.3781C} {461, 3781}

\bibitem[\protect\citeauthoryear{{Costa}, {Xu}, {Wang}  \& {Abdalla}}{{Costa}
  et~al.}{2017}]{Costa_2017JCAP}
{Costa} A.~A.,  {Xu} X.-D.,  {Wang} B.,   {Abdalla} E.,  2017, \mn@doi [Journal
  of Cosmology and Astro-Particle Physics] {10.1088/1475-7516/2017/01/028},
  \href {https://ui.adsabs.harvard.edu/abs/2017JCAP...01..028C} {2017, 028}

\bibitem[\protect\citeauthoryear{{DES Collaboration} et~al.,}{{DES
  Collaboration} et~al.}{2017}]{DES2017}
{DES Collaboration} et~al., 2017, ArXiv e-prints: 1708.01530, \href
  {http://adsabs.harvard.edu/abs/2017arXiv170801530D} {}

\bibitem[\protect\citeauthoryear{{Davis} \& {Peebles}}{{Davis} \&
  {Peebles}}{1983}]{Davis_1983ApJ}
{Davis} M.,  {Peebles} P.~J.~E.,  1983, \mn@doi [\apj] {10.1086/160884}, \href
  {https://ui.adsabs.harvard.edu/abs/1983ApJ...267..465D} {267, 465}

\bibitem[\protect\citeauthoryear{{Davis}, {Nusser}, {Masters}, {Springob},
  {Huchra}  \& {Lemson}}{{Davis} et~al.}{2011}]{davis2011}
{Davis} M.,  {Nusser} A.,  {Masters} K.~L.,  {Springob} C.,  {Huchra} J.~P.,
  {Lemson} G.,  2011, \mn@doi [\mnras] {10.1111/j.1365-2966.2011.18362.x},
  \href {https://ui.adsabs.harvard.edu/abs/2011MNRAS.413.2906D} {413, 2906}

\bibitem[\protect\citeauthoryear{Efron}{Efron}{1979}]{Efron_1979}
Efron B.,  1979, \mn@doi [Ann. Statist.] {10.1214/aos/1176344552}, 7, 1

\bibitem[\protect\citeauthoryear{{Feix}, {Nusser}  \& {Branchini}}{{Feix}
  et~al.}{2015}]{feix2015}
{Feix} M.,  {Nusser} A.,   {Branchini} E.,  2015, \mn@doi [\prl]
  {10.1103/PhysRevLett.115.011301}, \href
  {https://ui.adsabs.harvard.edu/abs/2015PhRvL.115a1301F} {115, 011301}

\bibitem[\protect\citeauthoryear{{Garc{\'{\i}}a-Farieta}, {Marulli},
  {Veropalumbo}, {Moscardini}, {Casas-Miranda}, {Giocoli}  \&
  {Baldi}}{{Garc{\'{\i}}a-Farieta} et~al.}{2019}]{garcia-farieta2019}
{Garc{\'{\i}}a-Farieta} J.~E.,  {Marulli} F.,  {Veropalumbo} A.,  {Moscardini}
  L.,  {Casas-Miranda} R.~A.,  {Giocoli} C.,   {Baldi} M.,  2019, \mn@doi
  [\mnras] {10.1093/mnras/stz1850}, \href
  {https://ui.adsabs.harvard.edu/abs/2019MNRAS.488.1987G} {488, 1987}

\bibitem[\protect\citeauthoryear{{Gil-Mar{\'\i}n}, {Wagner}, {Verde},
  {Porciani}  \& {Jimenez}}{{Gil-Mar{\'\i}n} et~al.}{2012}]{Gil_2012JCAP}
{Gil-Mar{\'\i}n} H.,  {Wagner} C.,  {Verde} L.,  {Porciani} C.,   {Jimenez} R.,
   2012, \mn@doi [Journal of Cosmology and Astro-Particle Physics]
  {10.1088/1475-7516/2012/11/029}, \href
  {https://ui.adsabs.harvard.edu/abs/2012JCAP...11..029G} {2012, 029}

\bibitem[\protect\citeauthoryear{{Granett}, {Favole}, {Montero-Dorta},
  {Branchini}, {Guzzo}  \& {de la Torre}}{{Granett}
  et~al.}{2019}]{Grannet_Multidark_2019}
{Granett} B.~R.,  {Favole} G.,  {Montero-Dorta} A.~D.,  {Branchini} E.,
  {Guzzo} L.,   {de la Torre} S.,  2019, arXiv e-prints, \href
  {https://ui.adsabs.harvard.edu/abs/2019arXiv190510375G} {p. arXiv:1905.10375}

\bibitem[\protect\citeauthoryear{{Guzzo} et~al.,}{{Guzzo}
  et~al.}{2000}]{guzzo2000}
{Guzzo} L.,  et~al., 2000, \aap, \href
  {http://adsabs.harvard.edu/abs/2000A%26A...355....1G} {355, 1}

\bibitem[\protect\citeauthoryear{{Guzzo} et~al.,}{{Guzzo}
  et~al.}{2014}]{Guzzo_VIPERS_2014}
{Guzzo} L.,  et~al., 2014, \mn@doi [\aap] {10.1051/0004-6361/201321489}, \href
  {https://ui.adsabs.harvard.edu/abs/2014A&A...566A.108G} {566, A108}

\bibitem[\protect\citeauthoryear{{Hamaus}, {Pisani}, {Sutter}, {Lavaux},
  {Escoffier}, {Wand elt}  \& {Weller}}{{Hamaus} et~al.}{2016}]{Hamaus2016}
{Hamaus} N.,  {Pisani} A.,  {Sutter} P.~M.,  {Lavaux} G.,  {Escoffier} S.,
  {Wand elt} B.~D.,   {Weller} J.,  2016, \mn@doi [\prl]
  {10.1103/PhysRevLett.117.091302}, \href
  {https://ui.adsabs.harvard.edu/abs/2016PhRvL.117i1302H} {117, 091302}

\bibitem[\protect\citeauthoryear{Hamilton}{Hamilton}{1998}]{Hamilton_review_1998}
Hamilton A. J.~S.,  1998, Linear Redshift Distortions: A Review.
Springer Netherlands, Dordrecht, pp 185--275

\bibitem[\protect\citeauthoryear{{Hartlap}, {Simon}  \& {Schneider}}{{Hartlap}
  et~al.}{2007}]{Hartlap2007}
{Hartlap} J.,  {Simon} P.,   {Schneider} P.,  2007, \mn@doi [\aap]
  {10.1051/0004-6361:20066170}, \href
  {https://ui.adsabs.harvard.edu/abs/2007A&A...464..399H} {464, 399}

\bibitem[\protect\citeauthoryear{{Hawken} et~al.,}{{Hawken}
  et~al.}{2017}]{hawken2017}
{Hawken} A.~J.,  et~al., 2017, \mn@doi [\aap] {10.1051/0004-6361/201629678},
  \href {https://ui.adsabs.harvard.edu/abs/2017A%26A...607A..54H} {607, A54}

\bibitem[\protect\citeauthoryear{{Howlett}, {Ross}, {Samushia}, {Percival}  \&
  {Manera}}{{Howlett} et~al.}{2015}]{howlett2015}
{Howlett} C.,  {Ross} A.~J.,  {Samushia} L.,  {Percival} W.~J.,   {Manera} M.,
  2015, \mn@doi [\mnras] {10.1093/mnras/stu2693}, \href
  {https://ui.adsabs.harvard.edu/abs/2015MNRAS.449..848H} {449, 848}

\bibitem[\protect\citeauthoryear{{Huterer}, {Shafer}, {Scolnic}  \&
  {Schmidt}}{{Huterer} et~al.}{2017}]{huterer2017}
{Huterer} D.,  {Shafer} D.~L.,  {Scolnic} D.~M.,   {Schmidt} F.,  2017, \mn@doi
  [\jcap] {10.1088/1475-7516/2017/05/015}, \href
  {https://ui.adsabs.harvard.edu/abs/2017JCAP...05..015H} {2017, 015}

\bibitem[\protect\citeauthoryear{{Ivezic} et~al.}{{Ivezic}
  et~al.}{2008}]{ivezic2008}
{Ivezic} Z.,  et~al., 2008, ArXiv e-prints: 0805.2366, \href
  {http://adsabs.harvard.edu/abs/2008arXiv0805.2366I} {}

\bibitem[\protect\citeauthoryear{Jackson}{Jackson}{1972}]{Jackson_FoG_1972}
Jackson J.~C.,  1972, \mn@doi [Monthly Notices of the Royal Astronomical
  Society] {10.1093/mnras/156.1.1P}, 156, 1P

\bibitem[\protect\citeauthoryear{{Jennings}}{{Jennings}}{2012}]{Jennings_2012}
{Jennings} E.,  2012, \mn@doi [\mnras] {10.1111/j.1745-3933.2012.01338.x},
  \href {https://ui.adsabs.harvard.edu/abs/2012MNRAS.427L..25J} {427, L25}

\bibitem[\protect\citeauthoryear{{Kaiser}}{{Kaiser}}{1987}]{Kaiser_1987}
{Kaiser} N.,  1987, \mn@doi [\mnras] {10.1093/mnras/227.1.1}, \href
  {http://adsabs.harvard.edu/abs/1987MNRAS.227....1K} {227, 1}

\bibitem[\protect\citeauthoryear{{Kazin}, {S{\'a}nchez}  \& {Blanton}}{{Kazin}
  et~al.}{2012}]{Kazin_2012_estimators}
{Kazin} E.~A.,  {S{\'a}nchez} A.~G.,   {Blanton} M.~R.,  2012, \mn@doi [\mnras]
  {10.1111/j.1365-2966.2011.19962.x}, \href
  {https://ui.adsabs.harvard.edu/abs/2012MNRAS.419.3223K} {419, 3223}

\bibitem[\protect\citeauthoryear{{Klypin}, {Yepes}, {Gottl{\"o}ber}, {Prada}
  \& {He{\ss}}}{{Klypin} et~al.}{2016}]{Klypin_Multidark_2016}
{Klypin} A.,  {Yepes} G.,  {Gottl{\"o}ber} S.,  {Prada} F.,   {He{\ss}} S.,
  2016, \mn@doi [\mnras] {10.1093/mnras/stw248}, \href
  {https://ui.adsabs.harvard.edu/abs/2016MNRAS.457.4340K} {457, 4340}

\bibitem[\protect\citeauthoryear{{Knebe} et~al.,}{{Knebe}
  et~al.}{2011}]{Halos_MAD_2011MNRAS}
{Knebe} A.,  et~al., 2011, \mn@doi [\mnras] {10.1111/j.1365-2966.2011.18858.x},
  \href {https://ui.adsabs.harvard.edu/abs/2011MNRAS.415.2293K} {415, 2293}

\bibitem[\protect\citeauthoryear{{Landy} \& {Szalay}}{{Landy} \&
  {Szalay}}{1993}]{LandySzalay_1993}
{Landy} S.~D.,  {Szalay} A.~S.,  1993, \mn@doi [\apj] {10.1086/172900}, \href
  {http://adsabs.harvard.edu/abs/1993ApJ...412...64L} {412, 64}

\bibitem[\protect\citeauthoryear{{Laureijs} et~al.,}{{Laureijs}
  et~al.}{2011}]{Laureijs_2011}
{Laureijs} R.,  et~al., 2011, arXiv e-prints, \href
  {https://ui.adsabs.harvard.edu/abs/2011arXiv1110.3193L} {p. arXiv:1110.3193}

\bibitem[\protect\citeauthoryear{Lewis, Challinor  \& Lasenby}{Lewis
  et~al.}{2000}]{Lewis_2000}
Lewis A.,  Challinor A.,   Lasenby A.,  2000, The Astrophysical Journal, 538,
  473

\bibitem[\protect\citeauthoryear{{Ling}, {Frenk}  \& {Barrow}}{{Ling}
  et~al.}{1986}]{Ling_1986}
{Ling} E.~N.,  {Frenk} C.~S.,   {Barrow} J.~D.,  1986, \mn@doi [\mnras]
  {10.1093/mnras/223.1.21P}, \href
  {http://adsabs.harvard.edu/abs/1986MNRAS.223P..21L} {223, 21P}

\bibitem[\protect\citeauthoryear{{Maartens}, {Abdalla}, {Jarvis}, {Santos}  \&
  {SKA Cosmology SWG}}{{Maartens} et~al.}{2015}]{Maartens_SKA_2015}
{Maartens} R.,  {Abdalla} F.~B.,  {Jarvis} M.,  {Santos} M.~G.,   {SKA
  Cosmology SWG} f.~t.,  2015, arXiv e-prints, \href
  {https://ui.adsabs.harvard.edu/abs/2015arXiv150104076M} {p. arXiv:1501.04076}

\bibitem[\protect\citeauthoryear{{Marulli}, {Carbone}, {Viel}, {Moscardini}  \&
  {Cimatti}}{{Marulli} et~al.}{2011}]{Marulli_2011MNRAS}
{Marulli} F.,  {Carbone} C.,  {Viel} M.,  {Moscardini} L.,   {Cimatti} A.,
  2011, \mn@doi [\mnras] {10.1111/j.1365-2966.2011.19488.x}, \href
  {https://ui.adsabs.harvard.edu/abs/2011MNRAS.418..346M} {418, 346}

\bibitem[\protect\citeauthoryear{{Marulli}, {Baldi}  \& {Moscardini}}{{Marulli}
  et~al.}{2012a}]{Marulli_2012MNRAS}
{Marulli} F.,  {Baldi} M.,   {Moscardini} L.,  2012a, \mn@doi [\mnras]
  {10.1111/j.1365-2966.2011.20199.x}, \href
  {https://ui.adsabs.harvard.edu/abs/2012MNRAS.420.2377M} {420, 2377}

\bibitem[\protect\citeauthoryear{{Marulli}, {Bianchi}, {Branchini}, {Guzzo},
  {Moscardini}  \& {Angulo}}{{Marulli}
  et~al.}{2012b}]{Marulli_anisotropies_2012MNRAS}
{Marulli} F.,  {Bianchi} D.,  {Branchini} E.,  {Guzzo} L.,  {Moscardini} L.,
  {Angulo} R.~E.,  2012b, \mn@doi [\mnras] {10.1111/j.1365-2966.2012.21875.x},
  \href {https://ui.adsabs.harvard.edu/abs/2012MNRAS.426.2566M} {426, 2566}

\bibitem[\protect\citeauthoryear{{Marulli}, {Veropalumbo}  \&
  {Moresco}}{{Marulli} et~al.}{2016}]{CosmoBolognaLib}
{Marulli} F.,  {Veropalumbo} A.,   {Moresco} M.,  2016, \mn@doi [Astronomy and
  Computing] {10.1016/j.ascom.2016.01.005}, \href
  {https://ui.adsabs.harvard.edu/abs/2016A%26C....14...35M} {14, 35}

\bibitem[\protect\citeauthoryear{{Marulli}, {Veropalumbo}, {Moscardini},
  {Cimatti}  \& {Dolag}}{{Marulli} et~al.}{2017}]{Marulli2017}
{Marulli} F.,  {Veropalumbo} A.,  {Moscardini} L.,  {Cimatti} A.,   {Dolag} K.,
   2017, \mn@doi [\aap] {10.1051/0004-6361/201526885}, \href
  {https://ui.adsabs.harvard.edu/abs/2017A%26A...599A.106M} {599, A106}

\bibitem[\protect\citeauthoryear{{Marulli} et~al.,}{{Marulli}
  et~al.}{2018}]{Marulli_XXL_2018}
{Marulli} F.,  et~al., 2018, \mn@doi [\aap] {10.1051/0004-6361/201833238},
  \href {https://ui.adsabs.harvard.edu/abs/2018A&A...620A...1M} {620, A1}

\bibitem[\protect\citeauthoryear{{Merloni} et~al.}{{Merloni}
  et~al.}{2012}]{Merloni2012}
{Merloni} A.,  et~al., 2012, ArXiv e-prints: 1209.3114, \href
  {http://adsabs.harvard.edu/abs/2012arXiv1209.3114M} {}

\bibitem[\protect\citeauthoryear{{Mohammad} et~al.,}{{Mohammad}
  et~al.}{2018}]{mohammad2018}
{Mohammad} F.~G.,  et~al., 2018, \mn@doi [\aap] {10.1051/0004-6361/201731685},
  \href {https://ui.adsabs.harvard.edu/abs/2018A%26A...610A..59M} {610, A59}

\bibitem[\protect\citeauthoryear{{Moresco} \& {Marulli}}{{Moresco} \&
  {Marulli}}{2017}]{Moresco2017}
{Moresco} M.,  {Marulli} F.,  2017, \mn@doi [\mnras] {10.1093/mnrasl/slx112},
  \href {https://ui.adsabs.harvard.edu/abs/2017MNRAS.471L..82M} {471, L82}

\bibitem[\protect\citeauthoryear{{Ntampaka}, {Rines}  \& {Trac}}{{Ntampaka}
  et~al.}{2019}]{Ntampaka_Multidark_2019}
{Ntampaka} M.,  {Rines} K.,   {Trac} H.,  2019, arXiv e-prints, \href
  {https://ui.adsabs.harvard.edu/abs/2019arXiv190607729N} {p. arXiv:1906.07729}

\bibitem[\protect\citeauthoryear{{Okumura} et~al.,}{{Okumura}
  et~al.}{2016}]{okumura2016}
{Okumura} T.,  et~al., 2016, \mn@doi [\pasj] {10.1093/pasj/psw029}, \href
  {https://ui.adsabs.harvard.edu/abs/2016PASJ...68...38O} {68, 38}

\bibitem[\protect\citeauthoryear{{Pacaud} et~al.,}{{Pacaud}
  et~al.}{2018}]{Pacaud_2018AA}
{Pacaud} F.,  et~al., 2018, \mn@doi [\aap] {10.1051/0004-6361/201834022}, \href
  {https://ui.adsabs.harvard.edu/abs/2018A&A...620A..10P} {620, A10}

\bibitem[\protect\citeauthoryear{{Parkinson} et~al.,}{{Parkinson}
  et~al.}{2012}]{WiggleZ_2012}
{Parkinson} D.,  et~al., 2012, \mn@doi [\prd] {10.1103/PhysRevD.86.103518},
  \href {https://ui.adsabs.harvard.edu/abs/2012PhRvD..86j3518P} {86, 103518}

\bibitem[\protect\citeauthoryear{{Peacock} \& {Dodds}}{{Peacock} \&
  {Dodds}}{1996}]{Peacock1996}
{Peacock} J.~A.,  {Dodds} S.~J.,  1996, \mnras, \href
  {http://cdsads.u-strasbg.fr/cgi-bin/nph-bib_query?bibcode=1996MNRAS.280L..19P&db_key=AST}
  {280, L19}

\bibitem[\protect\citeauthoryear{{Percival} et~al.}{{Percival}
  et~al.}{2004}]{percival2004}
{Percival} W.~J.,  et~al., 2004, \mn@doi [\mnras]
  {10.1111/j.1365-2966.2004.08146.x}, \href
  {http://adsabs.harvard.edu/abs/2004MNRAS.353.1201P} {353, 1201}

\bibitem[\protect\citeauthoryear{{Pezzotta} et~al.,}{{Pezzotta}
  et~al.}{2017}]{Pezzotta_vimos_2017}
{Pezzotta} A.,  et~al., 2017, \mn@doi [\aap] {10.1051/0004-6361/201630295},
  \href {https://ui.adsabs.harvard.edu/abs/2017A&A...604A..33P} {604, A33}

\bibitem[\protect\citeauthoryear{{Planck Collaboration} et~al.,}{{Planck
  Collaboration} et~al.}{2014}]{Planck_2014}
{Planck Collaboration} et~al., 2014, \mn@doi [\aap]
  {10.1051/0004-6361/201321591}, \href
  {https://ui.adsabs.harvard.edu/abs/2014A%26A...571A..16P} {571, A16}

\bibitem[\protect\citeauthoryear{{Planck Collaboration} et~al.,}{{Planck
  Collaboration} et~al.}{2016}]{Planck_2016}
{Planck Collaboration} et~al., 2016, \mn@doi [\aap]
  {10.1051/0004-6361/201525830}, \href
  {https://ui.adsabs.harvard.edu/abs/2016A&A...594A..13P} {594, A13}

\bibitem[\protect\citeauthoryear{{Planck Collaboration} et~al.,}{{Planck
  Collaboration} et~al.}{2018}]{Planck_Legacy_2018}
{Planck Collaboration} et~al., 2018, arXiv e-prints, \href
  {https://ui.adsabs.harvard.edu/abs/2018arXiv180706205P} {p. arXiv:1807.06205}

\bibitem[\protect\citeauthoryear{{Reid} et~al.}{{Reid} et~al.}{2012}]{reid2012}
{Reid} B.~A.,  et~al., 2012, \mn@doi [\mnras]
  {10.1111/j.1365-2966.2012.21779.x}, \href
  {http://adsabs.harvard.edu/abs/2012MNRAS.426.2719R} {426, 2719}

\bibitem[\protect\citeauthoryear{{Riebe} et~al.,}{{Riebe}
  et~al.}{2013}]{Riebe_2013AN_334}
{Riebe} K.,  et~al., 2013, \mn@doi [Astronomische Nachrichten]
  {10.1002/asna.201211900}, \href
  {https://ui.adsabs.harvard.edu/abs/2013AN....334..691R} {334, 691}

\bibitem[\protect\citeauthoryear{{Rodr{\'\i}guez-Puebla}, {Behroozi},
  {Primack}, {Klypin}, {Lee}  \& {Hellinger}}{{Rodr{\'\i}guez-Puebla}
  et~al.}{2016}]{Rodirguez_Multidark_2016}
{Rodr{\'\i}guez-Puebla} A.,  {Behroozi} P.,  {Primack} J.,  {Klypin} A.,  {Lee}
  C.,   {Hellinger} D.,  2016, \mn@doi [\mnras] {10.1093/mnras/stw1705}, \href
  {https://ui.adsabs.harvard.edu/abs/2016MNRAS.462..893R} {462, 893}

\bibitem[\protect\citeauthoryear{{Samushia}, {Percival}  \&
  {Raccanelli}}{{Samushia} et~al.}{2012}]{samushia2012}
{Samushia} L.,  {Percival} W.~J.,   {Raccanelli} A.,  2012, \mn@doi [\mnras]
  {10.1111/j.1365-2966.2011.20169.x}, \href
  {http://adsabs.harvard.edu/abs/2012MNRAS.420.2102S} {420, 2102}

\bibitem[\protect\citeauthoryear{{Samushia} et~al.,}{{Samushia}
  et~al.}{2014}]{samushia2014}
{Samushia} L.,  et~al., 2014, \mn@doi [\mnras] {10.1093/mnras/stu197}, \href
  {https://ui.adsabs.harvard.edu/abs/2014MNRAS.439.3504S} {439, 3504}

\bibitem[\protect\citeauthoryear{{S{\'a}nchez} et~al.,}{{S{\'a}nchez}
  et~al.}{2013}]{Sanchez2013}
{S{\'a}nchez} A.~G.,  et~al., 2013, \mn@doi [\mnras] {10.1093/mnras/stt799},
  \href {https://ui.adsabs.harvard.edu/abs/2013MNRAS.433.1202S} {433, 1202}

\bibitem[\protect\citeauthoryear{{S{\'a}nchez} et~al.,}{{S{\'a}nchez}
  et~al.}{2014}]{Sanchez2014}
{S{\'a}nchez} A.~G.,  et~al., 2014, \mn@doi [\mnras] {10.1093/mnras/stu342},
  \href {https://ui.adsabs.harvard.edu/abs/2014MNRAS.440.2692S} {440, 2692}

\bibitem[\protect\citeauthoryear{{S{\'a}nchez} et~al.,}{{S{\'a}nchez}
  et~al.}{2017}]{Sanchez2017}
{S{\'a}nchez} A.~G.,  et~al., 2017, \mn@doi [\mnras] {10.1093/mnras/stw2443},
  \href {https://ui.adsabs.harvard.edu/abs/2017MNRAS.464.1640S} {464, 1640}

\bibitem[\protect\citeauthoryear{{Santos} et~al.,}{{Santos}
  et~al.}{2015}]{Santos_SKA_2015}
{Santos} M.,  et~al., 2015, in Advancing Astrophysics with the Square Kilometre
  Array (AASKA14). p.~19 (\mn@eprint {arXiv} {1501.03989})

\bibitem[\protect\citeauthoryear{{Scoccimarro}}{{Scoccimarro}}{2004}]{Scoccimarro_model}
{Scoccimarro} R.,  2004, \mn@doi [\prd] {10.1103/PhysRevD.70.083007}, \href
  {https://ui.adsabs.harvard.edu/abs/2004PhRvD..70h3007S} {70, 083007}

\bibitem[\protect\citeauthoryear{{Sereno}, {Veropalumbo}, {Marulli}, {Covone},
  {Moscardini}  \& {Cimatti}}{{Sereno} et~al.}{2015}]{Sereno2015}
{Sereno} M.,  {Veropalumbo} A.,  {Marulli} F.,  {Covone} G.,  {Moscardini} L.,
   {Cimatti} A.,  2015, \mn@doi [\mnras] {10.1093/mnras/stv280}, \href
  {https://ui.adsabs.harvard.edu/abs/2015MNRAS.449.4147S} {449, 4147}

\bibitem[\protect\citeauthoryear{{Silk}}{{Silk}}{2017}]{Silk_2017nuco}
{Silk} J.,  2017, in 14th International Symposium on Nuclei in the Cosmos
  (NIC2016). p. 010101 (\mn@eprint {arXiv} {1611.09846}),
  \mn@doi{10.7566/JPSCP.14.010101}

\bibitem[\protect\citeauthoryear{Smith et~al.,}{Smith
  et~al.}{2003}]{Smith_halofit_2003}
Smith R.~E.,  et~al., 2003, \mn@doi [Monthly Notices of the Royal Astronomical
  Society] {10.1046/j.1365-8711.2003.06503.x}, 341, 1311

\bibitem[\protect\citeauthoryear{{Spergel} et~al.}{{Spergel}
  et~al.}{2013}]{Spergel2013}
{Spergel} D.,  et~al., 2013, ArXiv e-prints: 1305.5422, \href
  {http://adsabs.harvard.edu/abs/2013arXiv1305.5422S} {}

\bibitem[\protect\citeauthoryear{{Takahashi}, {Sato}, {Nishimichi}, {Taruya}
  \& {Oguri}}{{Takahashi} et~al.}{2012}]{Takahashi_2012ApJ}
{Takahashi} R.,  {Sato} M.,  {Nishimichi} T.,  {Taruya} A.,   {Oguri} M.,
  2012, \mn@doi [\apj] {10.1088/0004-637X/761/2/152}, \href
  {https://ui.adsabs.harvard.edu/abs/2012ApJ...761..152T} {761, 152}

\bibitem[\protect\citeauthoryear{{Taruya}, {Nishimichi}  \& {Saito}}{{Taruya}
  et~al.}{2010}]{TNS_model}
{Taruya} A.,  {Nishimichi} T.,   {Saito} S.,  2010, \mn@doi [\prd]
  {10.1103/PhysRevD.82.063522}, \href
  {https://ui.adsabs.harvard.edu/abs/2010PhRvD..82f3522T} {82, 063522}

\bibitem[\protect\citeauthoryear{{Tinker}, {Kravtsov}, {Klypin}, {Abazajian},
  {Warren}, {Yepes}, {Gottl{\"o}ber}  \& {Holz}}{{Tinker}
  et~al.}{2008}]{Tinker2008}
{Tinker} J.,  {Kravtsov} A.~V.,  {Klypin} A.,  {Abazajian} K.,  {Warren} M.,
  {Yepes} G.,  {Gottl{\"o}ber} S.,   {Holz} D.~E.,  2008, \mn@doi [\apj]
  {10.1086/591439}, \href
  {https://ui.adsabs.harvard.edu/abs/2008ApJ...688..709T} {688, 709}

\bibitem[\protect\citeauthoryear{{Tinker}, {Robertson}, {Kravtsov}, {Klypin},
  {Warren}, {Yepes}  \& {Gottl{\"o}ber}}{{Tinker}
  et~al.}{2010}]{Tinker_bias_2010}
{Tinker} J.~L.,  {Robertson} B.~E.,  {Kravtsov} A.~V.,  {Klypin} A.,  {Warren}
  M.~S.,  {Yepes} G.,   {Gottl{\"o}ber} S.,  2010, \mn@doi [\apj]
  {10.1088/0004-637X/724/2/878}, \href
  {http://adsabs.harvard.edu/abs/2010ApJ...724..878T} {724, 878}

\bibitem[\protect\citeauthoryear{{Tojeiro} et~al.}{{Tojeiro}
  et~al.}{2012}]{tojeiro2012}
{Tojeiro} R.,  et~al., 2012, \mn@doi [\mnras]
  {10.1111/j.1365-2966.2012.21404.x}, \href
  {http://adsabs.harvard.edu/abs/2012MNRAS.424.2339T} {424, 2339}

\bibitem[\protect\citeauthoryear{{Topping}, {Shapley}, {Steidel}, {Naoz}  \&
  {Primack}}{{Topping} et~al.}{2018}]{Topping_Multidark_2018}
{Topping} M.~W.,  {Shapley} A.~E.,  {Steidel} C.~C.,  {Naoz} S.,   {Primack}
  J.~R.,  2018, \mn@doi [\apj] {10.3847/1538-4357/aa9f0f}, \href
  {https://ui.adsabs.harvard.edu/abs/2018ApJ...852..134T} {852, 134}

\bibitem[\protect\citeauthoryear{{Upadhye}}{{Upadhye}}{2019}]{Upadhye_2019JCAP}
{Upadhye} A.,  2019, \mn@doi [Journal of Cosmology and Astro-Particle Physics]
  {10.1088/1475-7516/2019/05/041}, \href
  {https://ui.adsabs.harvard.edu/abs/2019JCAP...05..041U} {2019, 041}

\bibitem[\protect\citeauthoryear{{Vega-Ferrero}, {Yepes}  \&
  {Gottl{\"o}ber}}{{Vega-Ferrero} et~al.}{2017}]{Vega-Ferrero_Multidark_2017}
{Vega-Ferrero} J.,  {Yepes} G.,   {Gottl{\"o}ber} S.,  2017, \mn@doi [\mnras]
  {10.1093/mnras/stx282}, \href
  {https://ui.adsabs.harvard.edu/abs/2017MNRAS.467.3226V} {467, 3226}

\bibitem[\protect\citeauthoryear{{Wang} et~al.,}{{Wang}
  et~al.}{2018}]{Wang_Multidark_2018}
{Wang} Y.,  et~al., 2018, \mn@doi [\apj] {10.3847/1538-4357/aae52e}, \href
  {https://ui.adsabs.harvard.edu/abs/2018ApJ...868..130W} {868, 130}

\bibitem[\protect\citeauthoryear{{Wright}, {Koyama}, {Winther}  \&
  {Zhao}}{{Wright} et~al.}{2019}]{Wright_2019JCAP}
{Wright} B.~S.,  {Koyama} K.,  {Winther} H.~A.,   {Zhao} G.-B.,  2019, \mn@doi
  [Journal of Cosmology and Astro-Particle Physics]
  {10.1088/1475-7516/2019/06/040}, \href
  {https://ui.adsabs.harvard.edu/abs/2019JCAP...06..040W} {2019, 040}

\bibitem[\protect\citeauthoryear{Xu, Padmanabhan, Eisenstein, Mehta  \&
  Cuesta}{Xu et~al.}{2012}]{Xu_2012A}
Xu X.,  Padmanabhan N.,  Eisenstein D.~J.,  Mehta K.~T.,   Cuesta A.~J.,  2012,
  \mn@doi [Monthly Notices of the Royal Astronomical Society]
  {10.1111/j.1365-2966.2012.21573.x}, 427, 2146

\bibitem[\protect\citeauthoryear{Xu, Cuesta, Padmanabhan, Eisenstein  \&
  McBride}{Xu et~al.}{2013}]{Xu_2012B}
Xu X.,  Cuesta A.~J.,  Padmanabhan N.,  Eisenstein D.~J.,   McBride C.~K.,
  2013, \mn@doi [Monthly Notices of the Royal Astronomical Society]
  {10.1093/mnras/stt379}, 431, 2834

\bibitem[\protect\citeauthoryear{{Zandanel}, {Fornasa}, {Prada}, {Reiprich},
  {Pacaud}  \& {Klypin}}{{Zandanel} et~al.}{2018}]{Zandanel_Multidark_2018}
{Zandanel} F.,  {Fornasa} M.,  {Prada} F.,  {Reiprich} T.~H.,  {Pacaud} F.,
  {Klypin} A.,  2018, \mn@doi [\mnras] {10.1093/mnras/sty1901}, \href
  {https://ui.adsabs.harvard.edu/abs/2018MNRAS.480..987Z} {480, 987}

\bibitem[\protect\citeauthoryear{{Zheng}, {Zhang}  \& {Oh}}{{Zheng}
  et~al.}{2017}]{Zheng_2017JCAP}
{Zheng} Y.,  {Zhang} P.,   {Oh} M.,  2017, \mn@doi [Journal of Cosmology and
  Astro-Particle Physics] {10.1088/1475-7516/2017/05/030}, \href
  {https://ui.adsabs.harvard.edu/abs/2017JCAP...05..030Z} {2017, 030}

\bibitem[\protect\citeauthoryear{{de la Torre} \& {Guzzo}}{{de la Torre} \&
  {Guzzo}}{2012}]{delaTorre_2012MNRAS}
{de la Torre} S.,  {Guzzo} L.,  2012, \mn@doi [\mnras]
  {10.1111/j.1365-2966.2012.21824.x}, \href
  {https://ui.adsabs.harvard.edu/abs/2012MNRAS.427..327D} {427, 327}

\bibitem[\protect\citeauthoryear{{de la Torre} et~al.}{{de la Torre}
  et~al.}{2013}]{delatorre2013b}
{de la Torre} S.,  et~al., 2013, \mn@doi [\aap] {10.1051/0004-6361/201321463},
  \href {http://adsabs.harvard.edu/abs/2013A%26A...557A..54D} {557, A54}

\bibitem[\protect\citeauthoryear{{de la Torre} et~al.,}{{de la Torre}
  et~al.}{2017}]{delatorre2017}
{de la Torre} S.,  et~al., 2017, \mn@doi [\aap] {10.1051/0004-6361/201630276},
  \href {https://ui.adsabs.harvard.edu/abs/2017A%26A...608A..44D} {608, A44}

\bibitem[\protect\citeauthoryear{{van den Bosch} \& {Jiang}}{{van den Bosch} \&
  {Jiang}}{2016}]{vandenBosh_Multidark_2014}
{van den Bosch} F.~C.,  {Jiang} F.,  2016, \mn@doi [\mnras]
  {10.1093/mnras/stw440}, \href
  {https://ui.adsabs.harvard.edu/abs/2016MNRAS.458.2870V} {458, 2870}

\makeatother
\end{thebibliography}

\bsp	
\label{lastpage}
\end{document}